\documentclass[conference,10pt]{IEEEtran}
\usepackage{amsmath,amssymb,amsfonts}
\usepackage{float}
\usepackage{algorithmic}
\usepackage{graphicx}
\usepackage{textcomp}
\usepackage{xcolor}
\newcommand{\etal}{\textit{et al.}}
\newcommand{\ourtit}{NNReArch}
\usepackage{hyphenat}
\usepackage{caption}
\usepackage{subcaption}
\usepackage{tablefootnote}
\usepackage[numbers,sort&compress]{natbib}
\usepackage{url}

\begin{document}

\title{\huge{\ourtit: A Tensor Program Scheduling Framework Against Neural Network Architecture Reverse Engineering}}

\author{\IEEEauthorblockN{Yukui Luo, Shijin Duan, Cheng Gongye, Yunsi Fei, and Xiaolin Xu}
\IEEEauthorblockA{{Department of Electrical and Computer Engineering}}
\IEEEauthorblockA{{Northeastern University, Boston, MA, USA}}}

\maketitle

\begin{abstract}
Architecture reverse engineering has become an emerging attack against deep neural network (DNN) implementations. Several prior works have utilized side-channel leakage to recover the model architecture while the target is executing on a hardware acceleration platform. In this work, we target an open-source deep-learning accelerator, Versatile Tensor Accelerator (VTA), and utilize electromagnetic (EM) side-channel leakage to comprehensively learn the association between DNN architecture configurations and EM emanations. We also consider the holistic system -- including the low-level tensor program code of the VTA accelerator on a Xilinx FPGA, and explore the effect of such low-level configurations on the EM leakage. Our study demonstrates that both the optimization and configuration of tensor programs will affect the EM side-channel leakage. 

Gaining knowledge of the association between the low-level tensor program and the EM emanations, we propose \ourtit, a lightweight tensor program scheduling framework against side-channel-based DNN model architecture reverse engineering. Specifically, NNReArch targets reshaping the EM traces of different DNN operators, through scheduling the tensor program execution of the DNN model so as to confuse the adversary. NNReArch is a comprehensive protection framework supporting two modes, a \textit{balanced mode} that strikes a balance between the DNN model confidentiality and execution performance, and a \textit{secure mode} where the most secure setting is chosen. We implement and evaluate the proposed framework on the open-source VTA with state-of-the-art DNN architectures. The experimental results demonstrate that NNReArch can efficiently enhance the model architecture security with a small performance overhead. In addition, the proposed obfuscation technique makes reverse engineering of the DNN architecture significantly harder.

\end{abstract}

\section{Introduction}\label{sec:inro}
Neural network (NN) has found important use in different application domains, such as object detection, big data analytic, and semantic recognition. To improve the inference capability of NN models, deep neural network (DNN) is proposed, which employs larger model size for tackling more complicated tasks. Although DNNs are demonstrated as promising for different tasks, their large model sizes become a bottleneck for performance and applicability. To mitigate these issues, different methods have been proposed to accelerate the execution of DNN models \cite{jouppi2017datacenter,Xilinx_DPU,chen2018tvm,chen2016eyeriss, luo2016dadiannao}. For example, %field-programmable gate array (FPGA) has become one of the most important hardware acceleration platforms of DNN \cite{Xilinx_DPU}. Taking the advantages of customized hardware implementation and software instruction set architecture, 
modern FPGAs have been widely deployed to provide acceleration for DNNs on both edge devices and cloud infrastructures. 

A DNN model can be represented as a directed acyclic graph (DAG) composed of operation nodes and connections, thus its implementation can be mapped accordingly to FPGA hardware components (e.g., look-up-table and DSP) with FPGA-DNN development tools. The most commonly used frameworks are the Xilinx deep learning processor unit (DPU)~\cite{Xilinx_DPU}, and the open-source versatile tensor accelerator (VTA)~\cite{moreau2018vta}, which can provide end-to-end optimization for FPGA-DNN implementation. Although such FPGA-based DNN acceleration framework provide significant performance improvement, they also create a new attack surface, where an adversary can either manipulate the inference of DNN models \cite{luo2021deepstrike,rakin2021deep}, or illegally extract (i.e., using side-channel analysis) the critical parameters of a DNN model, such as its architecture \cite{yu2020deepem,zhou2021deep}. Since the construction of high-performance DNN models involves expensive data collection and training procedures, thus their model parameters like the architecture should be highly protected.

This paper studies the vulnerability of DNN model architecture against side-channel-based DNN model extraction attack on FPGA accelerators. Specifically, we explore the association between  electromagnetic (EM) side-channel leakage of DNN model implementation on FPGAs. To draw generic conclusions, we use state-of-the-art open-source FPGA acceleration framework, VTA \cite{VTA_PYNQ_Z1}, and explore the internal causes of DNN model architecture-relevant side-channel leakage from the low-level program code.  Correspondingly, we propose defense solutions by rescheduling the DNN operator-level tensor program and obfuscating the side-channel leakage. Note that although this paper mainly discusses EM side-channel and VTA, the presented experimental observation and the proposed methodologies can be generally extended to other side-channels and hardware platforms. 

The main contributions of this work are as follows:

\begin{itemize}
    \item We comprehensively study the EM side-channel leakage for DNN models deployed with VTA. To the best of our knowledge, this is the first work exploring the association between low-level tensor program code and EM trace, from the adversarial perspective.
    
    \item We systematically formulate the mathematical representation of the DNN architecture with EM emanations, following which we further propose security metrics for the DNN models based on the visualized EM characteristics. These metrics can be used to estimate the security level of a DNN architecture against side-channel attacks.
    
    \item We present \ourtit, a flexible defense framework for DNN model architectures against EM side-channel attacks. We evaluate the performance of \ourtit~with state-of-the-art DNN architectures, and we demonstrate that \ourtit~can significantly complex the reverse engineering attacks, i.e., doubling the attacking efforts with only 3.06\% performance overhead. %, as the VGG-19 example shows.

\end{itemize}

%The remainder of this paper is organized as follows: Sec.~\ref{sec:main} presents the EM side-channel leakage and corresponding low-level tensor program code. Sec.~\ref{sec:nnrearch_sch} details the proposed defense framework \ourtit, and the experimental evaluations are presented in Sec.~\ref{sec:eval}. Sec.~\ref{sec:discussion}  concludes this work.

\section{Background and Related work}\label{sec:bg}

\subsection{Versatile Tensor Accelerator (VTA)}\label{sec:bg-vta}

VTA is an open-source, generic, and  FPGA-specific deep learning acceleration framework~\cite{moreau2018vta}, consisting of four modules to enable task-level parallelism in a pipelining fashion (TLPP): a ``fetch module'' that loads the instruction stream from the DRAM, a ``load module'' loading the input, weight parameters, and intermediate results, a ``store module''  writing back intermediate results, and the ``compute module'' accelerates computing. The last one is ``compute module" that relies on two kernels, the general matrix multiply (GEMM) kernel for dense linear algebra computations, and the tensor arithmetic logic unit (ALU) for general computing tasks. VTA is supported by TVM~\cite{chen2018tvm}, a compiler (AutoTVM) scheduling the target deep learning application on a hardware platform. In addition, two techniques, matrix multiplication blocking (MMB) and virtual threading (VT), are used to customize the FPGA execution to further improve the VTA performance. The MMB technology divides large NN operators down to smaller blocks to fit the GEMM kernel, and VT manages the hardware resources of VTA to facilitate simultaneous computing and memory access. More technical details of the VTA can be found in Sec.~\ref{sec:main}.

\subsection{Model Architecture Extraction Attack and Defense}

Model architecture extraction has become an emerging threat to the security of DNN. In addition to attacks from the software side~\cite{jagielski2020high}, different side-channels have been utilized to extract DNN architectures on hardware platforms. Batina \etal~\cite{batina2019csi} first demonstrated architecture extraction of multi-layer perception (MLP) using the EM signals from both AVR and ARM processors. In~\cite{yu2020deepem}, Yu \etal~applied EM-based attack on an FPGA against a convolutional neural network (CNN).  Tian \etal~\cite{tian2021remote} utilized an on-chip time-to-digital (TDC) sensor to extract the NN architecture, which can be launched remotely on a multi-tenant FPGA, but with lower resolutions compared to EM signals. Hu \etal~\cite{hu2020deepsniffer} demonstrated extracting a complete NN architecture using multiple side-channels of GPUs, such as the memory access pattern. However, none of the prior works targets comprehensively studying the tensor programmable accelerator, such as VTA. Also, defense methods are still in their infancy. %A recent work~\cite{li2021neurobfuscator} introduces a graph-based obfuscation method against DNN model extraction attacks, which targets at mis-leading the model-parameter inference by modifying the internal DNN connection and architecture. 

\subsection{EM Side-channel}
EM side-channel \cite{batina2019csi} is an effective and contact-less method for extracting sensitive information during the system execution. The instantaneous EM emanation is dependent on the dynamic current  \cite{yu2020deepem}, as shown in Eq. \ref{eq:EM_dynamic}: 
\begin{equation}\label{eq:EM_dynamic}
\small
    I_{dyn}(t) = \frac{C\times V_{DD}\times f_{clk}\times D(t)}{2} 
\end{equation} 
where $C$ denotes the capacitance of the activated metal nets, $D(t)$ represents the transition rate of the nets (determined by both operations and data), $V_{DD}$ is the voltage supply, and $f_{clk}$ stands for the execution clock frequency \cite{anderson2004power}. Therefore, an EM trace from a hardware platform well embodies information for its workload, data, and system computing and communication. 

\section{Associating Tensor Program on VTA with EM Emanation}\label{sec:main}

\begin{figure}[t!]
  \centering
  \includegraphics[width=0.7\linewidth]{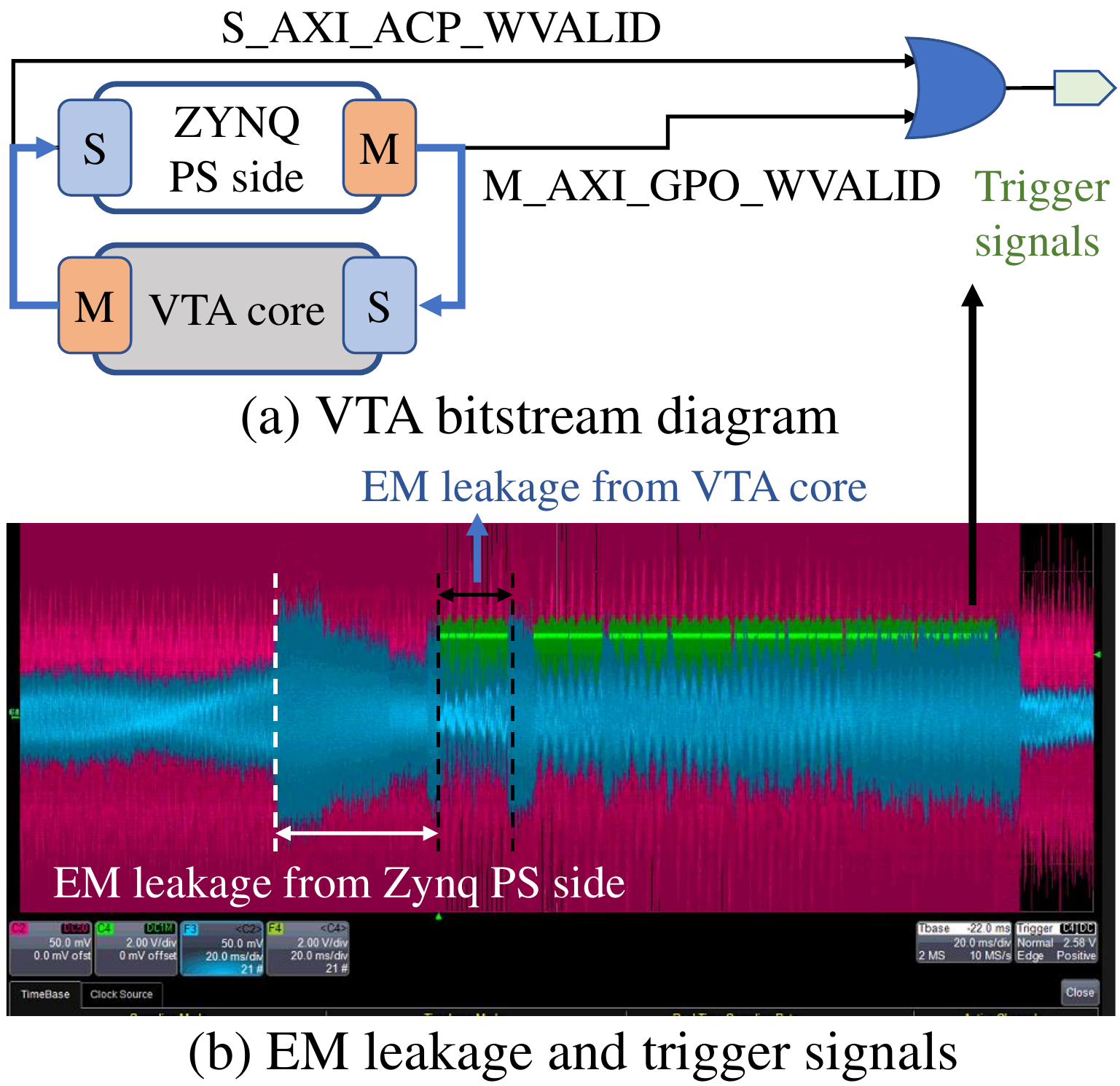}
  \caption{Experimental Platform for EM Signal Collection} %diagram. (a) shows the experimental VTA bitstream, which triggers the EM measurement system. (b) demonstrates the EM trace of a NN executed on our experiment platform. We note that the specific scheduler (a.k.a AutoTVM) for the VTA assigns some computing on the Zynq PS side, and our trigger can annotate the computing on the VTA core.}
  \label{fig:VTA_platform}
  \vspace{-1.5em}
\end{figure}

\subsection{Threat model}
We follow the same threat model as in other related side-channel model extraction works \cite{yu2020deepem, batina2019csi}, with our victim device being an edge FPGA running an VTA accelerator for a pre-trained DNN model. The attacker is able to obtain the EM signals of the victim device with specialized equipment and also knows the model execution status, which can assist in reasoning the architecture of the victim DNN model. We assume a strong attacker, who has sufficient knowledge of the target accelerator,  including the configuration of the accelerator (discussed in Sec.~\ref{sec:vta-config}). The executing model architecture is the target for reverse engineering with all the EM traces, accelerator, and platform information. 

\if false 
that demonstrate NN architecture extraction attack using EM side-channel, as detailed below. 
\begin{enumerate}
    \item We consider the victim devices as edge FPGA accelerators developed by VTA, on which the model provider deploys NN models. The DNN model includes the architecture and parameters (e.g., weight and basis), which are pre-loaded and ready for inference. %NN applications can implement with a hardware accelerator. Therefore, 
    Note that  our proposed protection methodology and analysis can also be applied to other hardware accelerators with a program scheduler.
    
    \item Attackers have physical access to the local device, which means they have the ability and time to explore the machine thoroughly. Specifically, they can use an electromagnetic (EM) probe to trace the EM leakage repeatedly, including the model execution status, which can assist attackers in reasoning the architecture of the victim DNN model.
    
    \item Our threat model further considers a stronger assumption that, the attackers have sufficient knowledge of the target accelerator, i.e., they know the configuration of the victim accelerator (as discussed in Sec.~\ref{sec:vta-config}). These configurations can assist attackers in analyzing the EM leakage and to accurately extract the DNN architecture.
\end{enumerate}
\fi 

\begin{figure*}[htpb!]
     \centering
     \begin{subfigure}[b]{0.32\textwidth}
         \centering
         \includegraphics[width=\textwidth]{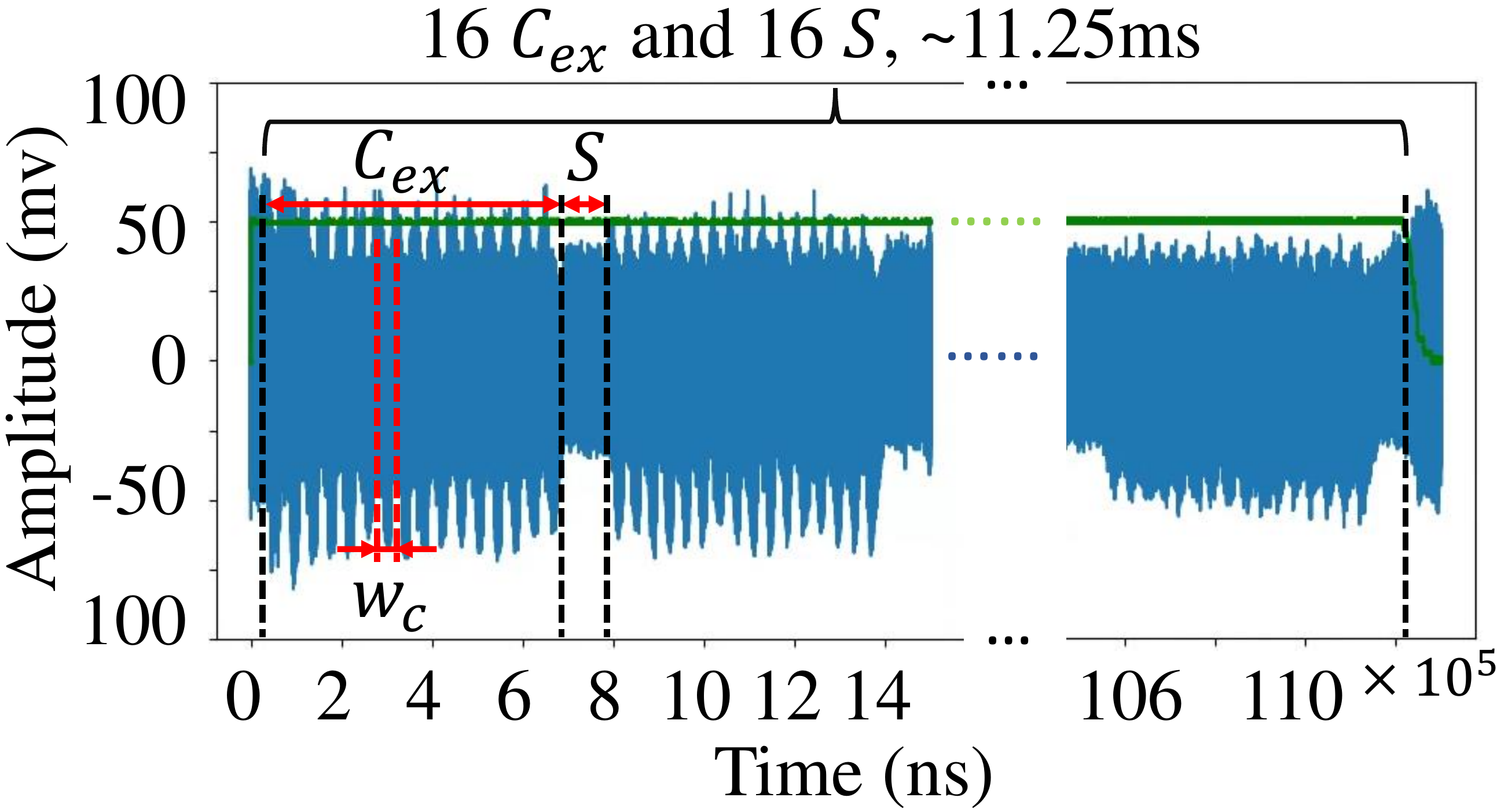}
         \caption{Conv2D-Opt:[1, 1, 1, 1, 1]. The baseline w/o optimization.}
         \label{fig:Wop-ex-baseline}
     \end{subfigure}
     \hfill
     \begin{subfigure}[b]{0.32\textwidth}
         \centering
         \includegraphics[width=\textwidth]{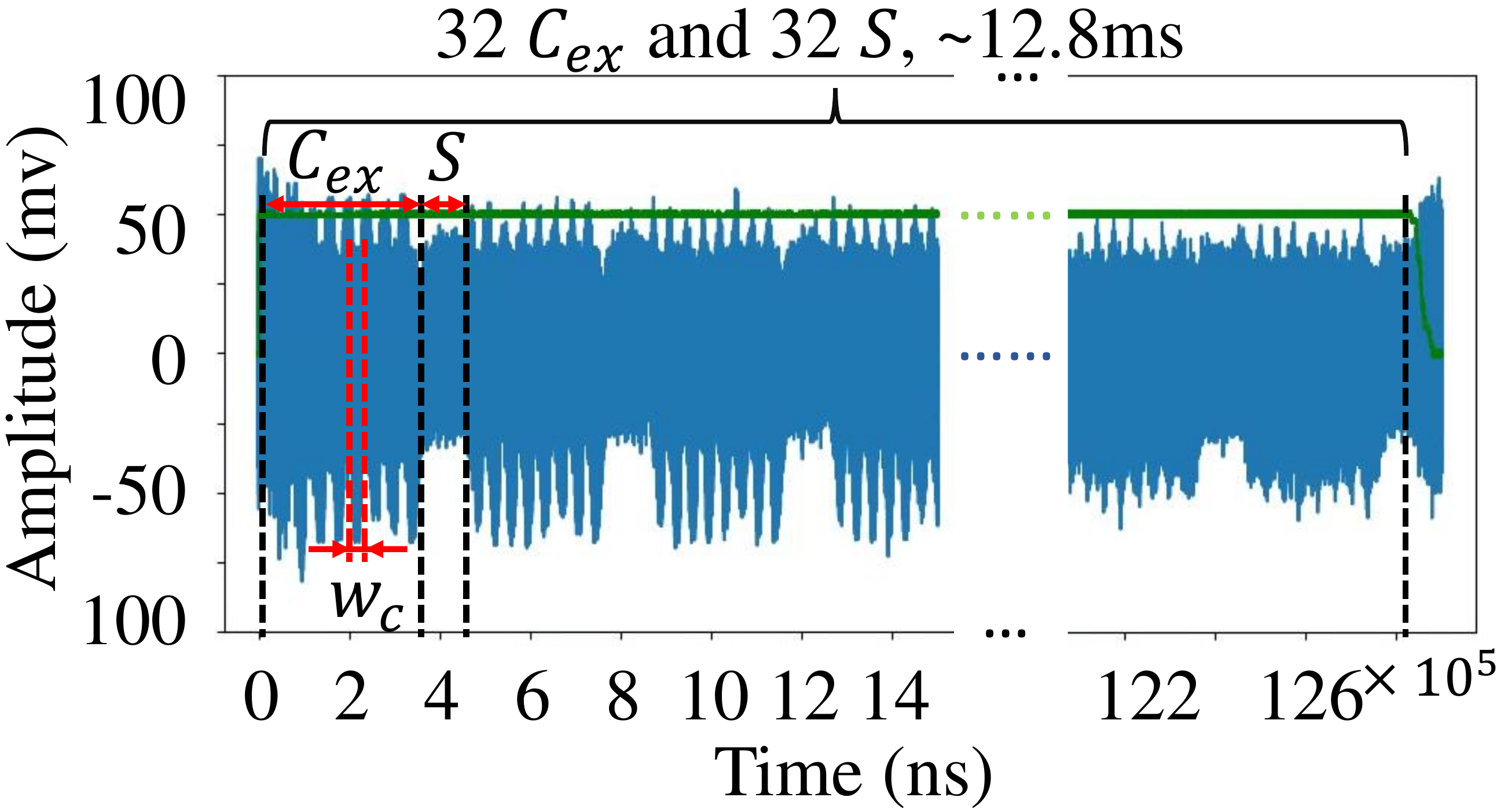}
         \caption{Conv2D-Opt:[2, 1, 1, 1, 1]. Blocks along the input channel ($ic_b=2$).}
         \label{fig:Wop-ex-ic}
     \end{subfigure}
     \hfill
     \begin{subfigure}[b]{0.32\textwidth}
         \centering
         \includegraphics[width=\textwidth]{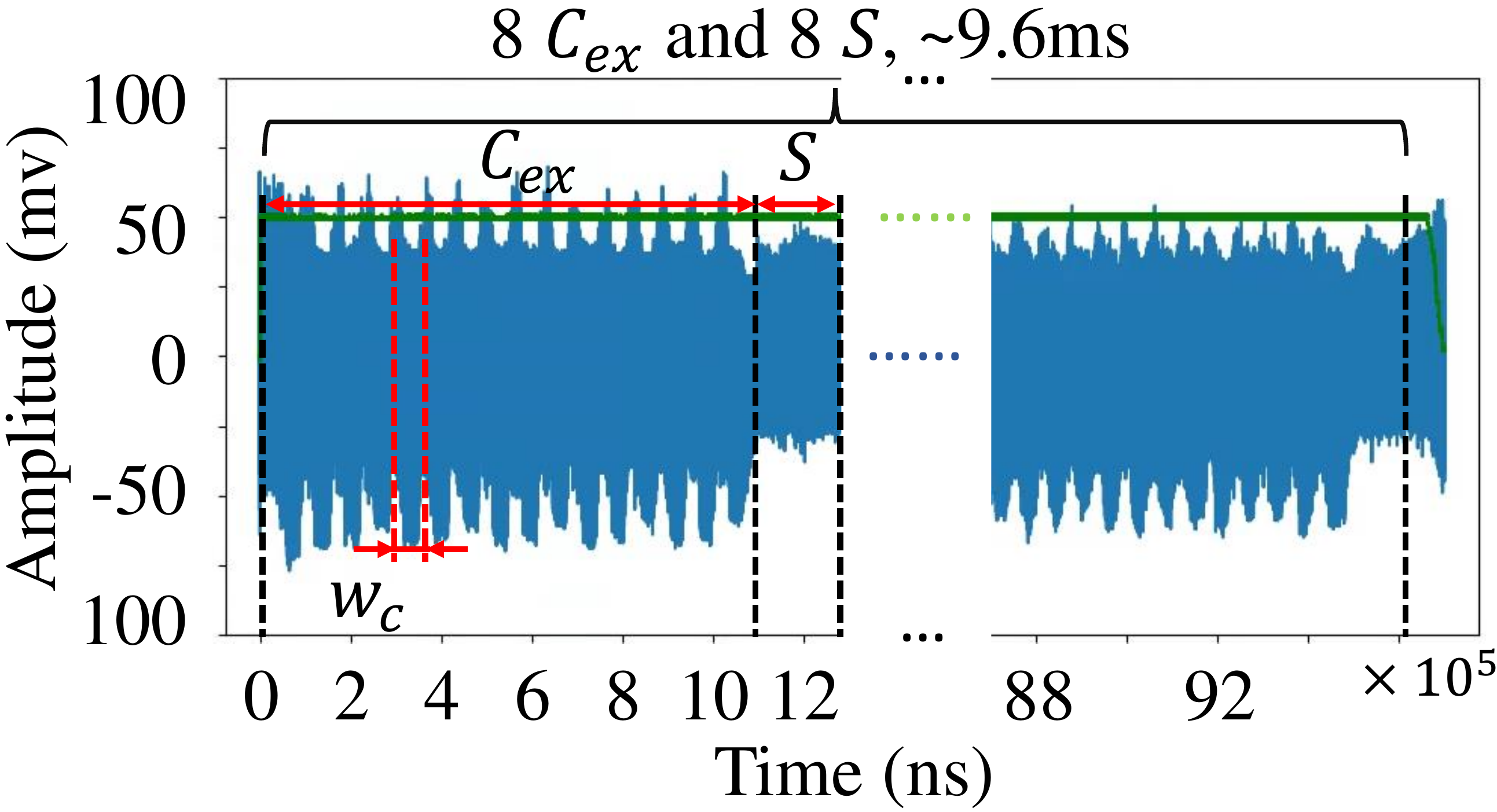}
         \caption{Conv2D-Opt:[1, 2, 1, 1, 1]. Blocks along the output channel ($oc_b=2$).}
         \label{fig:Wop-ex-oc}
     \end{subfigure}
     \begin{subfigure}[b]{0.32\textwidth}
         \centering
         \includegraphics[width=\textwidth]{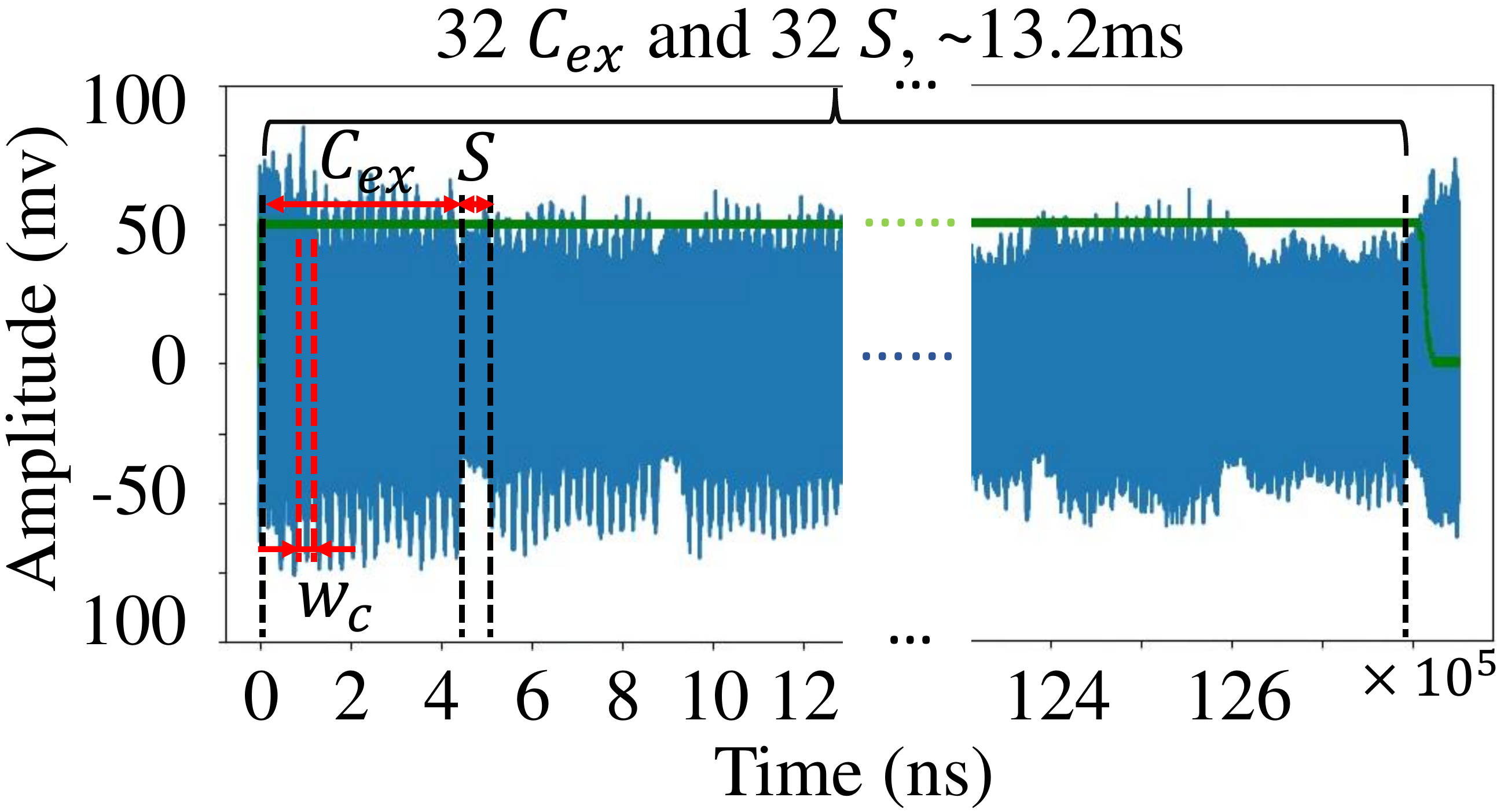}
         \caption{Conv2D-Opt:[1, 1, 2, 1, 1]. Blocks along the input feature height ($fih_b=2$).}
         \label{fig:Wop-ex-fih}
     \end{subfigure}
     \hfill
     \begin{subfigure}[b]{0.32\textwidth}
         \centering
         \includegraphics[width=\textwidth]{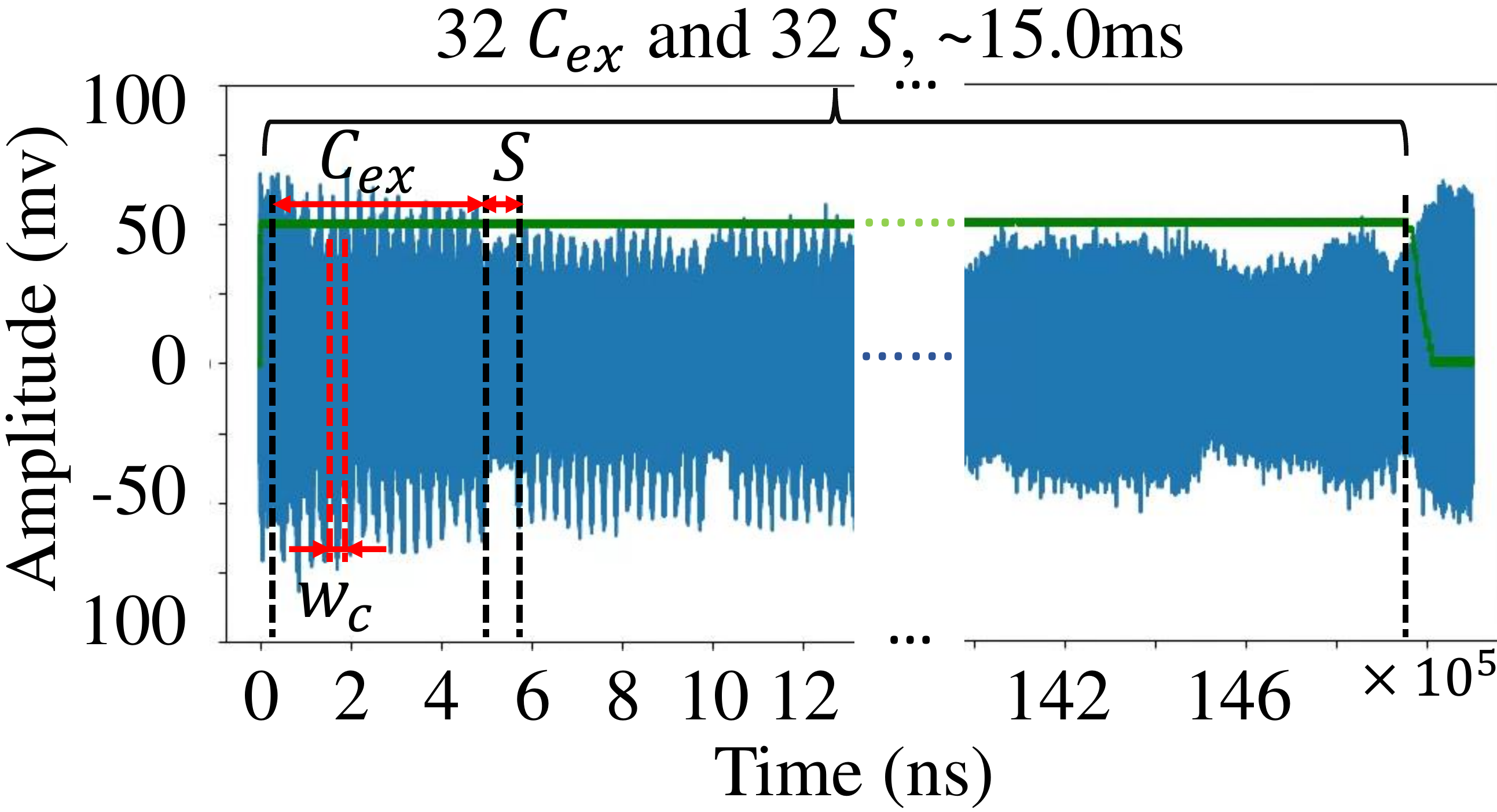}
         \caption{Conv2D-Opt:[1, 1, 1, 2, 1]. Blocks  along the input feature width ($fiw_b=2$).}
         \label{fig:Wop-ex-fiw}
     \end{subfigure}
     \hfill
     \begin{subfigure}[b]{0.32\textwidth}
         \centering
         \includegraphics[width=\textwidth]{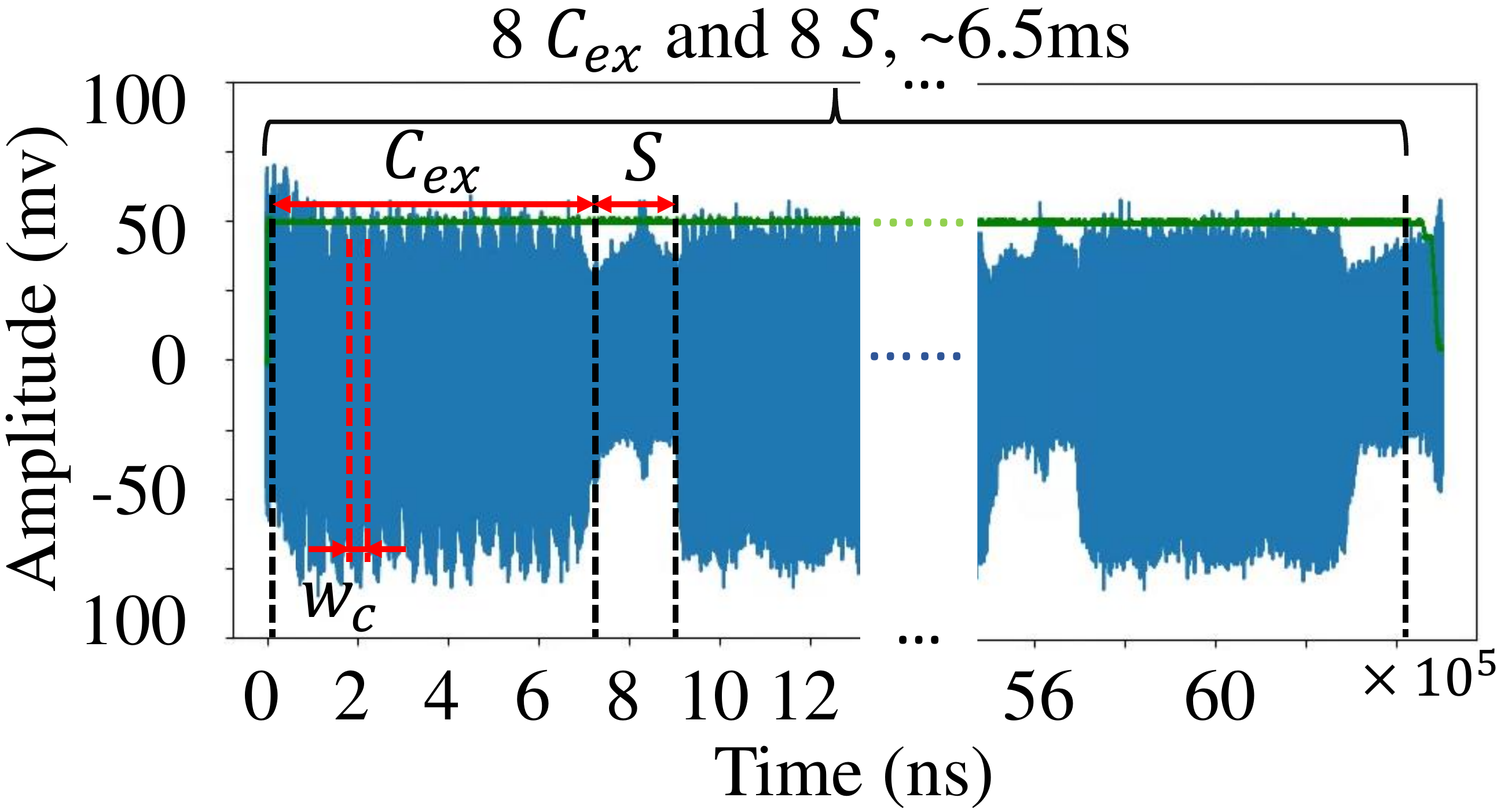}
         \caption{Conv2D-Opt:[1, 1, 1, 1, 2]. Apply the virtual threading ($vt=2$).}
         \label{fig:Wop-ex-vt}
     \end{subfigure}
        \caption{We use Conv2D-Wop:[256, 256, 3, 14, 14] as an example to show how the EM leakages change with the Opt-config [$ic_b, oc_b, fih_b, fiw_b, vt$].}
        \label{fig:Wop-ex}
        \vspace{-1em}
\end{figure*}

\subsection{Experimental Platform}

We build an experimental platform using PYNQ-Z1 development kit, an SoC with a Xilinx Zynq-7000 device and a dual-core ARM Cortex-A9 processor (PS). To align the EM signals with execution phases for the device characterization purpose, we modify the VTA bitstream as shown in Fig.\ref{fig:VTA_platform} (a). Specifically, we use two signals as the trigger signals to mark the starting and ending of the VTA execution (and therefore the corresponding EM segment): the general-purpose output (GPO) manager write valid of the PS, \textit{M\_AXI\_GPO\_WVALID}, and the accelerator coherence port (ACP) subordinate write valid of the VTA core, \textit{S\_AXI\_ACP\_WVALID}. The \textit{M\_AXI\_GPO\_WVALID} signal annotates the opcodes and data that have been written into the VTA queues, and \textit{S\_AXI\_ACP\_WVALID} indicates that the VTA core output is valid to be written back to the DRAM of the PYNQ-Z1 system. 

Our EM trace collection setup includes an EM Probe PBS2~\cite{AaroniaP68:online} converting the EM signals into voltage representations, an Aronia AG pre-amplifier, and a Lecroy oscilloscope~\cite{Teledyne19:online}. An EM trace example with the sampling rate of 1GHz is shown in Fig. \ref{fig:VTA_platform} (b), which is averaged from 50 measurements with the same inputs. The average pre-processing method helps to make the EM trace stable and filter the measurement noise. 

\subsection{Terminology and Definitions}
With the EM measurement setup fixed, there are three other factors that jointly determine the EM trace measurement of a VTA: (1) The workload configuration of the current DNN layer's operation, namely \textbf{Wop-config}; (2) The global configuration of the VTA-core, namely \textbf{VTA-config}; (3) To take advantage of the FPGA parallelism and ARM multi-thread scheduling, operators (resources) for a DNN layer are optimized, whose configuration is denoted as \textbf{Opt-config}. 

\subsubsection{Wop-config} For the most commonly used DNN architectures, 2D convolutional layer (Conv2D) is the most critical component to construct the entire model architecture. A Conv2D is specified by the number of input channels ($IC$), output channels ($OC$), the kernel size ($K$), the input feature size ($FI$), and the output feature size ($FO$). We follow the typical regulation that assumes the input/output feature is square-shaped. The Conv2D-Wop is therefore [$IC, OC, K, FI, FO$].
%For example, a Conv2D-Wop of [$256,256,3,14,14$] represents a Conv2D layer with 256 input channels and 256 output channels. The filter kernel size is $3 \times 3$, and the input feature map size is $14 \times 14$. With 1-pixel spatial padding and 1-pixel stride scanning, the output feature map size is $14 \times 14$ as well. %Note that the padding size is $\frac{K-1}{2}$.%, and the scanning stride is $\frac{FO}{FI}, FO\leq FI$. If $FO > FI$, it will 

\subsubsection{VTA-config}\label{sec:vta-config} A VTA-core is deployed on an FPGA specified by VTA-config, the configuration from~\cite{VTA_PYNQ_Z1}. The main computing component GEMM kernel, is designed around a tensor core performing one matrix-matrix operation in each clock cycle. This operator is to implement the product of a $1 \times 16$ input and a $16 \times 16$ weight matrix. The VTA core employs hardware resources for parallel computation to achieve high performance. The input matrix has dimension of $BATCH\times BLOCK\_IN$, where $BATCH$ indicates how many feature maps can be implemented in parallel, by the VTA core ($BATCH$ = 1 by default), and $BLOCK\_IN$ represents the input channel-parallelism. For example, in our experimental, $BLOCK\_IN$ is set as 16, indicating that 16 input channels can execute in parallel. The weight matrix includes $BLOCK\_IN$ $\times$ $BLOCK\_OUT$ number of weights, where $BLOCK\_OUT$ represents the output channel-parallelism. When $BLOCK\_OUT$ is 16, the VTA can produce results in 16 output channels (output $BATCH \times  BLOCK\_OUT$). Another setup of VTA-config is the sizes of on-chip buffers, including input, output, and weight buffer with 32KB, 128KB, and 32KB memory sizes, respectively.

\subsubsection{Opt-config} Rather than static execution, the VTA can dynamically schedule the execution of Conv2D layers for performance optimization. AutoTVM supports VTA to implement explicit memory latency hiding by the virtual threading ($vt$) primitive, corresponding to multi-threading of the ARM processor. As our used ARM Cortex A9 dual-core processor allows two threads, the VTA $vt$ can support threads up to 2. To map the matrix multiplication efficiently on a VTA core, TVM can optimally break down large workload as smaller blocks, to achieve computation efficiency within limited hardware resources. There are four scales of blocks associated with this technology: %batch blocks ($b_b$), 
input channel blocks ($ic_b$), output channel blocks ($oc_b$), and two input feature map blocks along the height axis ($fih_b$) and width axis ($fiw_b$), respectively. We use an Opt-config vector $[ic_b, oc_b, fih_b, fiw_b, vt]$ to represent the optimization setting. For example, a  Conv2D-Opt of [2, 2, 2, 2, 2] is applied on a convolution layer with Con2D-Wop: [256, 256, 3, 14, 14]. The scheduler will divide the original convolution layer into several small blocks with workload $Conv2D_b$-Wop:[128, 128, 3, 14, 14] because both the input ($ic_b$) and output channel ($oc_b$) blocks are 2. It will also separate the $14 \times 14$ input feature map into $7 \times 7$ blocks because $fih_b$ and $fiw_b$ are also 2. Moreover, the Opt-config enables dual-threading. In contrast, the non-optimized Opt-config is Conv2D-Opt:[1, 1, 1, 1, 1]. Note that across this paper, we use the default $BATCH$ setting, and the subscript $_b$ is used to represent the detailed value of each parameter out of many possibilities.

\subsection{EM Leakage Observation}\label{sec:observe}
In our experiment, we implement a convolutional layer with Conv2D-Wop of [256, 256, 3, 14, 14]. We choose different Opt-configs to understand the impact of optimizations on DNN execution, which is reflected in the EM leakage. Fig.~\ref{fig:Wop-ex} shows the EM traces collected from the basic VTA setting without optimization (Fig.~\ref{fig:Wop-ex-baseline}) and five optimized versions (Fig.~\ref{fig:Wop-ex-ic} to \ref{fig:Wop-ex-vt}). Inspecting these traces, we find a repetitive pattern of a segment of high-frequency activity (continuous execution, $C_{ex}$) followed by a segment of low-frequency (stalling) activity ($S$). This pattern repeats $M$ times for the convolutional layer computation. 
%In one Conv2D, we assume that stalls separate the trace into $M$ parts, each of which implements continuous matrix multiplication. We use $C_{ex}$ to annotate its continuous execution time. As shown in Fig.~\ref{fig:Wop-ex}, 
Further, from the beginning $C_{ex}$ segment, we can clearly observe several spikes, %\footnote{Paper \cite{gnad2017voltage} discussed the reason why those $N$ spikes are blurred after the Conv2D executes for a while.}.
the number ($N$) of which is countable and each of them has approximately the same width ($w_c$), where $C_{ex} = N \times w_c$. Thus, we can derive a Conv2D EM trace function ($Conv2D_{EM}$) with 2 countable parameters: $M$ and $N$, and 2 measurable parameters: $w_c$ and $S$.
\vspace{-0.5em}
\begin{equation}\label{eq:Conv2D_EM}
\small
    Conv2D_{EM} = M \times (N \times w_c + S)
\end{equation}

\begin{figure}[t!]
  \centering
  \includegraphics[width=\linewidth]{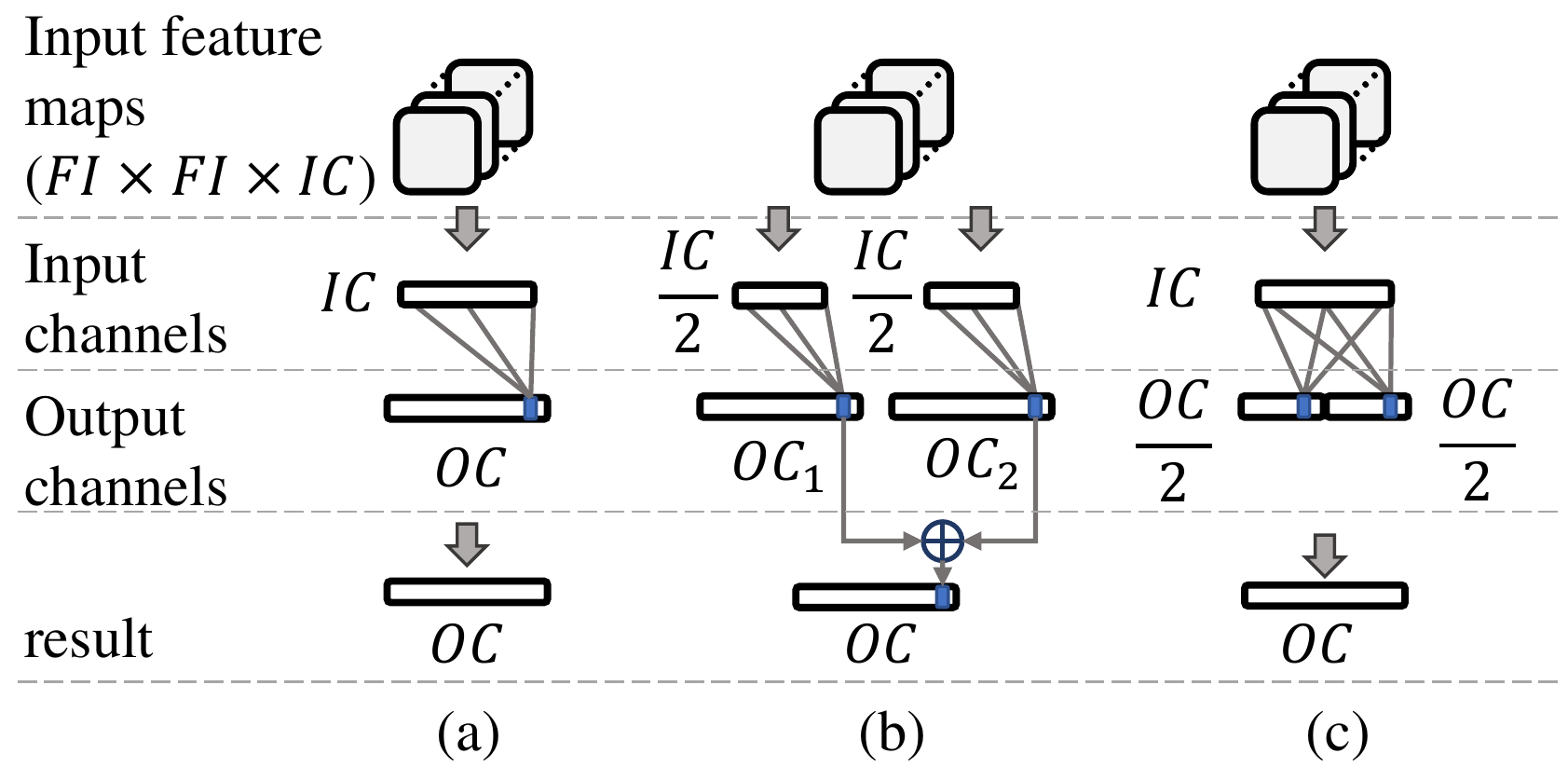}
  \caption{The scheduler divides a large Conv2D layer down to smaller blocks along the IC and OC axes. (a) w/o optimization. (b)  Conv2D-Opt: [2, 1, 1, 1, 1], $ic_b = 2$. (c) Conv2D-Opt: [1, 2, 1, 1, 1],  $oc_b = 2$.}
% \Description{.}
  \label{fig:IC_OC_B}
  \vspace{-1em}
\end{figure}

When we change the $ic_b$ along the $IC$ and $oc_b$ along the $OC$ of the Opt-config, the computing flow for each axis' blocks is shown in Fig.~\ref{fig:IC_OC_B}, and their corresponding EM traces are shown in Fig.~\ref{fig:Wop-ex-ic} and Fig~\ref{fig:Wop-ex-oc}. If dividing the Conv2D layer along the $IC$ axis, it will generate $ic_b$ subordinated outputs $OC_{i}$, whose summation will derive the final result. When $ic_b = 2$, two paths are scheduled and the execution time of $C_{ex}$ is reduced by almost a half. Blocks along the $OC$ axis has a simpler computational process. The scheduler separates $OC$ output channels into $oc_b$ sections, and the final result is the concatenation of those sub-results. Comparing Fig.~\ref{fig:Wop-ex-baseline} with Fig.~\ref{fig:Wop-ex-ic} and Fig.~\ref{fig:Wop-ex-oc}, the $IC$ axis blocking determines $M$ and $N$, and the $OC$ axis blocking determines $M$ and $w_c$. Similarly, blocks along the feature map height and width also affect $M$ and $w_c$ of the EM trace, as shown in Fig.~\ref{fig:Wop-ex-fih} and Fig.~\ref{fig:Wop-ex-fiw}, although no obvious difference between these two EM traces can be observed. Following our measurement results, blocking along the width of the feature map ($fiw_b$) induces a longer $S$.
%In terms of the input feature map arrangement, we can observe the changes on $M$ and $w_c$ in Fig.~\ref{fig:Wop-ex-fih} and Fig.~\ref{fig:Wop-ex-fiw}, compared with the baseline. They apply the blocking method along height and weight axes of the input feature map separately. However, no obvious difference between Fig.~\ref{fig:Wop-ex-fih} and Fig.~\ref{fig:Wop-ex-fiw} can be observed.
% we can observe a slight execution time difference between Fig.~\ref{fig:Wop-ex-fih} and Fig.~\ref{fig:Wop-ex-fiw}. They apply the blocking method alone input feature maps' height and weight axes separately, and they change $M$ and $w_c$ compare with the baseline. 
Besides, we applied the virtual threading method, which accelerates the operator by hiding the DRAM memory access latency and enables the TLPP of VTA as mentioned in Sec.\ref{sec:bg-vta}. As shown in Fig.~\ref{fig:Wop-ex-vt}, this configuration shortens the execution time of the entire Conv2D layer compared with the baseline by reducing the $M$. 

From the EM leakage observation, we can draw the following conclusions: (1) $M$ is a function of $IC$, $ic_b$, $OC$, $oc_b$, $FO$, $FI$, $fih_b$, $fiw_b$, and $vt$; (2) $N$ is a function of $IC$, $ic_b$; and  (3) $w_c$ is a function of $OC$, $oc_b$, $FI$, $fih_b$, and $fiw_b$.

\begin{figure*}
    \centering
  \includegraphics[width=0.95\linewidth]{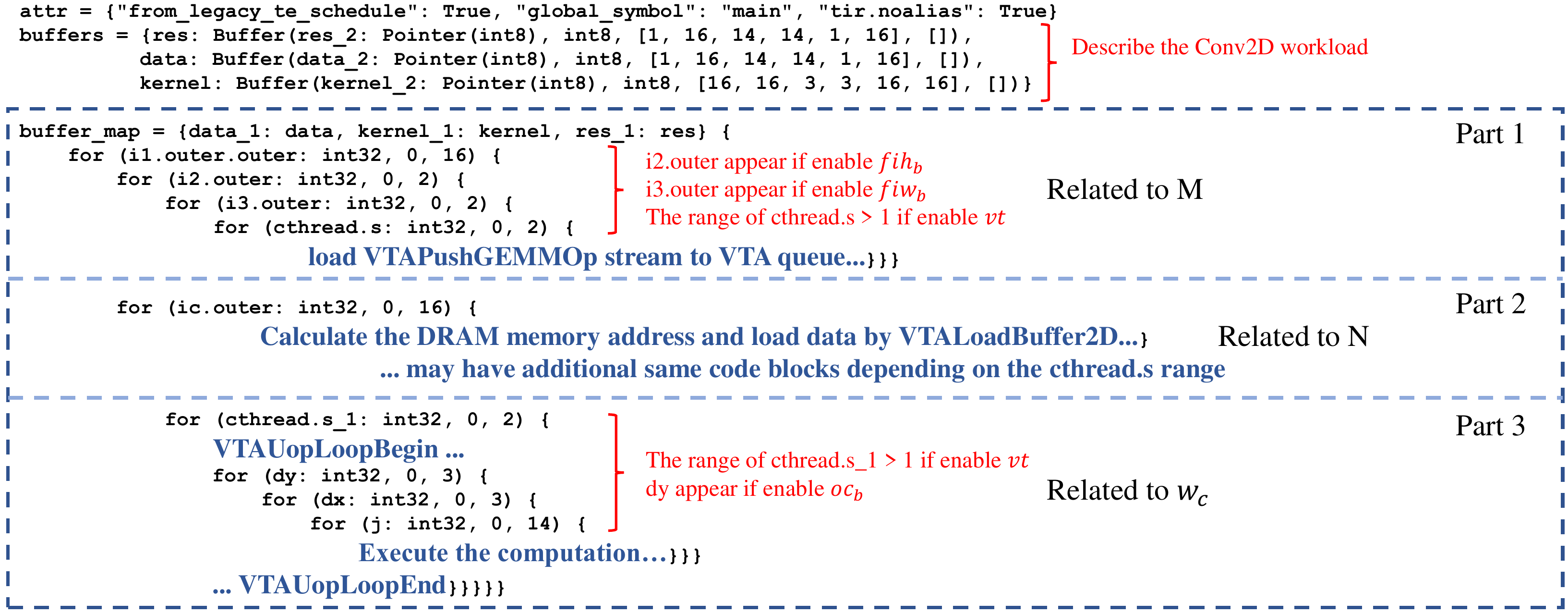}
  \caption{Low-level code summary for the optimizable and high-arithmetic intensity GEMM operator, which constructs the Conv2D layer with low-arithmetic intensity ALU.}
  \label{fig:con2d_lower_code}
  \vspace{-1.5em}
\end{figure*}

\subsection{Low-level Program Code Analysis}\label{sec:analysis}
Visually inspecting the EM traces derives general association between the EM pattern and the two configurations, Wop-config and Opt-config. To comprehensively understand the execution impacts on the EM leakage, we look into the low-level code structure shown in Fig~\ref{fig:con2d_lower_code}. %Fig~\ref{fig:Wop-ex}.
% We can learn more if we construct the EM leakage with the operator's low-level code. 
% We summarised the low-level code of the above Conv2D layer EM leakage observation in Fig~\ref{fig:Wop-ex}. 
Since the Tensor ALU operators of a Conv2D layer have low-arithmetic intensity and therefore do not emanate high EM leakage, we focus on the GEMM operator~\cite{moreau2018vta}. %Thus, We focus on its high-arithmetic intensity GEMM operator optimization analysis, as shown in Fig.~\ref{fig:con2d_lower_code}. % As Fig.~\ref{fig:con2d_lower_code} shows, t 
The GEMM code is composed by many nested loops of operations, corresponding to the repetitive EM pattern shown in Fig~\ref{fig:Wop-ex}. We extract three parts (part 1 to 3 as shown in Fig~\ref{fig:con2d_lower_code}) related to the $Conv2D_{EM}$ function parameters, $M$, $N$, and $w_c$, respectively. In Part 1, there are four outer loops related to $M$, and their ranges indicate the blocking parameters: $ic_b$, $fih_b$, $fiw_b$ and $vt$. Note that the part 1 program is the most outer loop, the function of $M$ is also determined by several other parameters, as defined in Eq.~\ref{eq:M}:
%the range of \textit{i1.outer.outer} is related to $OC$, $BLOCK\_OUT$, $oc_b$, and $ic_b$, the range of \textit{i2.outer} or \textit{i3.outer} is equivalent to $fih_b$ or $fiw_b$, and the range of \textit{cthread.s} is equal to $vt$ %larger than 1, if the virtual threading optimization is enabled. %, and its range is equal to $vt$. Thus, we can calculate
\vspace{-0.3em}
\begin{equation}\label{eq:M}
\small
    M=\frac{IC \times ic_b \times FO \times fih_b \times fiw_b}{BLOCK\_OUT \times FI \times oc_b \times vt}
\end{equation}

It is straightforward to determine $N$ from the range of \texttt{ic.outer} of the Part 2 code: 
%, as shown in Fig.~\ref{fig:con2d_lower_code}, with the following function: 
\vspace{-0.3em}
\begin{equation}\label{eq:N}
\small
    N = \frac{IC}{BLOCK\_IN \times ic_b}
\end{equation}

% Part 3 is related to $w_c$. 
Different from $M$ and $N$ that are discrete (integer) numbers, $w_c$ is associated with the execution time. Hence, without knowing the exact function, we can only leverage the Part 3 code to determine which parameters affect its quantity. In the first \texttt{cthread.s\_1} loop, if its range is larger than 1, it will enable the TLPP. Our experiments suggest that the range of \texttt{dx} is equal to $K$, i.e., the kernel size. If $oc_{b} = 1$, the range of \texttt{dy} is also 1, otherwise it is $K$. The range of $j$ is a function of $FI$, $fiw_b$, and $FO$.
%since $\frac{FI}{FO}$ is defined as the scanning stride in the Conv2D layer.
%, and its normal value is 1 or 2, we can obtain $FO$ with a reasonable assumption of stride \red{(I'm not sure about this)}. 
Putting all these clues together, we assume function $g(\cdot)$ can obtain $w_c$ from low-level tensor program code in Eq.~\ref{eq:wc}.
\vspace{-0.3em}
\begin{equation}\label{eq:wc}
\small
    w_c = g(K, FI, fiw_b, FO, vt, f_{ex}, II)
\end{equation}

The TLPP is configured by $vt$, $II$ denotes the initiation interval for the pipeline, and $f_{ex}$ is the executing frequency of the VTA core, which is $100MHz$ in this paper. 

\subsection{EM Obfuscation}\label{sec:EMob}

\begin{figure}[htpb!]
     \centering
          \begin{subfigure}[b]{\linewidth}
         \centering
         \includegraphics[width=0.9\linewidth]{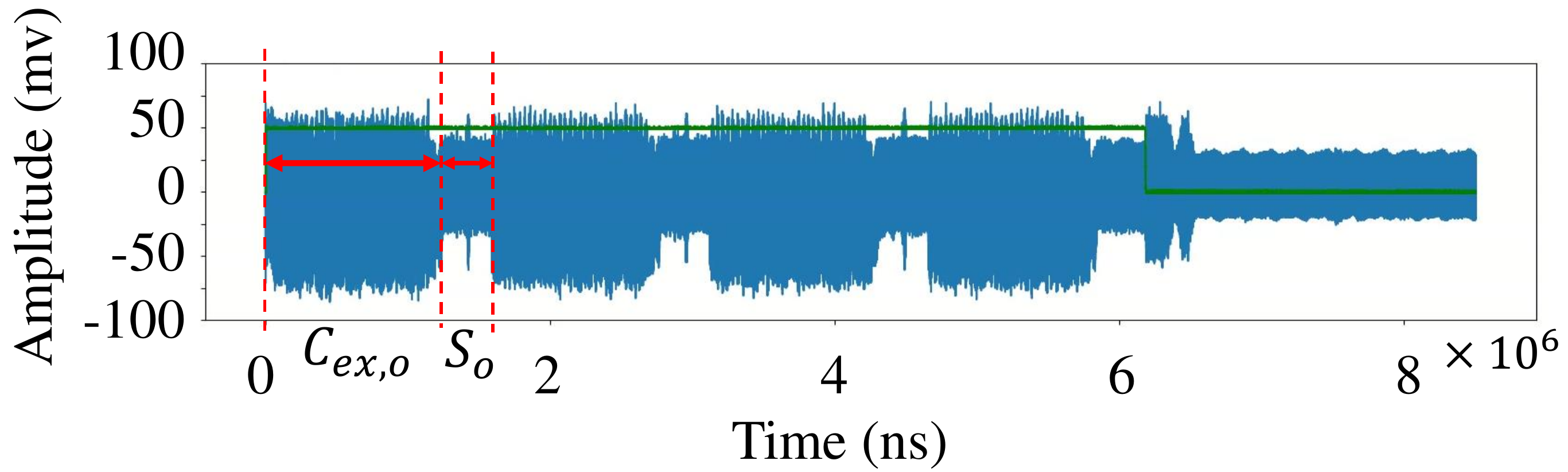}
         \caption{Conv2D$_{o}$-Wop:[256, 256, 3, 14, 14], Conv2D$_{o}$-Opt:[1, 2, 1, 1, 2].}
         \label{fig:obf_org}
     \end{subfigure}
     \begin{subfigure}[b]{\linewidth}
         \centering
         \includegraphics[width=0.9\linewidth]{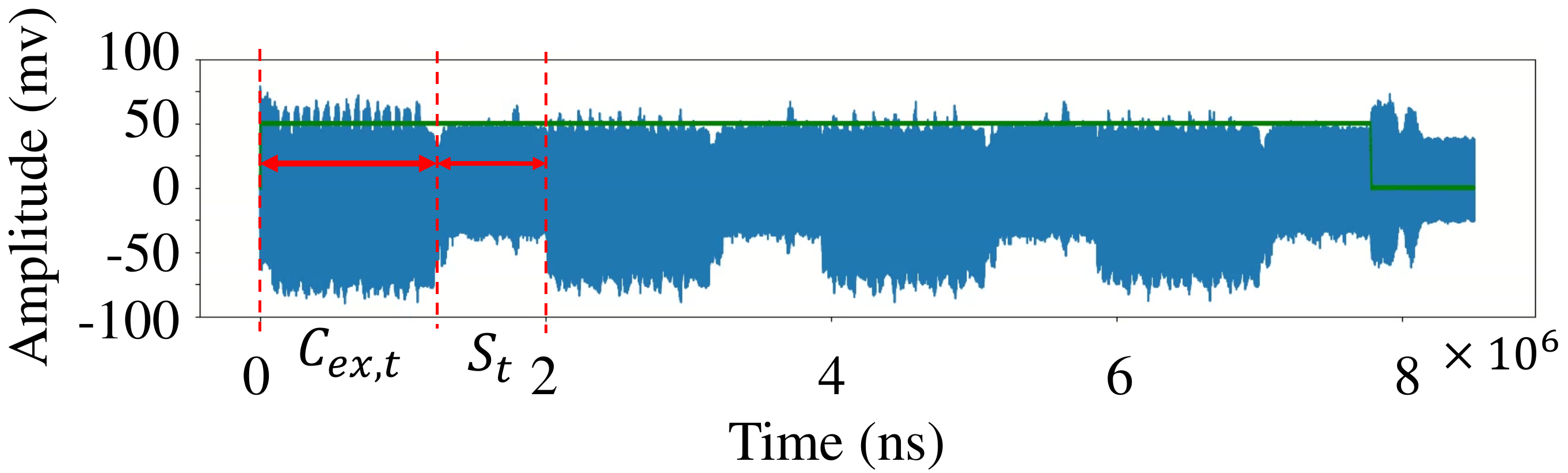}
         \caption{Conv2D$_{t}$-Wop:[128, 128, 3, 28, 28], Conv2D$_{t}$-Opt:[1, 4, 2, 2, 2].}
         \label{fig:obf_tag}
     \end{subfigure}

        \caption{Conv2D EM obfuscation example. We use the subscript $o$ to represent the original Conv2D layer, and subscript $t$ is the target obfuscation Conv2D layer.}
        \label{fig:obf}
        \vspace{-1em}
\end{figure}

With the EM leakage characterization and the low-level code analysis, we propose to obfuscate the EM trace by scheduling the tensor program. As a proof of concept, we implement two different Conv2D layers: one Conv2D$_{o}$ with the Wop-config of [256, 256, 3, 14, 14] and Opt-config of[1, 2, 1, 1, 2]; the other Conv2D$_{t}$ with the Wop-config of [128, 128, 3, 28, 28] and Opt-config of [1, 4, 2, 2, 2]. Inspecting  their EM traces in Fig. \ref{fig:obf}, we notice these two layers can generate similar $C_{ex}$, because their GEMM operators have equal counting result ($OpC_{GEMM}$) derived by Eq.~\ref{eq:pcgemm}.
\vspace{-0.3em}
\begin{equation}\label{eq:pcgemm}
\small
    OpC_{GEMM} = \frac{IC \times OC \times K^2 \times FO^2}{BLOCK\_IN \times BLOCK\_OUT}
\end{equation}

However, the stall time is different between these two settings ($S_{o}$ and $S_{t}$). Such difference can be canceled by adding \textbf{pause opcode} to delay $\sim 0.35ms$ in every $S_{o}$, so that $S_{o}'=S_{t}$. As a result, these two EM traces become in-distinguishable( i.e., unable to determine which setting is in effect). 

Generally, the goal of EM obfuscation for a Conv2D layer is to find different configurations resulting in the same operation counts as the original one, following Eq.~\ref{eq:pcgemm}. Specifically, this equation can assist us to find a target Wop-config $Conv2D_{t}-Wop$, for which there exists an Opt-config $Conv2D_{t}-Opt$ satisfying $M_t = M_o$, $K_o = K_t$, and has a longer execution time. Then we can derive $\Delta S = S_{t} - S_{o}$ by measurement, and apply it to the original workload to mimic it as the target workload.

% \section{Generic NN architecture extraction attack and secure metrics}

\section{Security Metrics and Optimization Program for VTA Implementation}\label{sec:nnrearch_sch}
This section introduces %the proposed FPGA-DNN architecture protection framework, 
\ourtit, which utilizes Opt-config and EM obfuscation to mitigate the EM leakage of the DNN model. For an attacker to reverse-engineer the victim NN architecture implemented on VTA, s/he needs to derive VTA-config and Opt-config. If Opt-config is fixed, such as Conv2D-Opt of [1, 1, 1, 1, 1], then the victim architecture is easy to extract. Thus, a primary idea of mitigating the EM side-channel leakage is to increase  the searching space of the Opt-config for each Conv2D layer.

A designer can formulate the scheduling of DNN execution as an optimization problem. For a given neural network architecture $NN$, we can extract a set of workload expression $E$ that executes on a target acceleration device. Then, for a given workload $e \in E$, we can implement it %through many different executions 
with many different functionally equivalent low-level program codes inducing different EM traces, as observed in Sec.~\ref{sec:observe} and~\ref{sec:analysis}. Therefore, each workload could have multiple equivalent schedules, i.e., Opt-config. We use $Ps_{e}$ to denote the possible schedule space for $e$. For example, in VGG-19, there are 9 types of Conv2D layers with different Wop-configs, each of whic is denoted as $e_{i}, i \in [1, 9]$ and has a set of Opt-config $Ps_{e_i}$.

\subsection{Security Metrics}
When considering confidentiality of a $NN$, brute force attack is the most generic method and its complexity can be represented by the size of its searching space ($SS_{NN}$ defined in Eq. \ref{eq:ssnn}). The attacker normally progresses sequentially, i.e., from the first Conv2D layer to the following layers, since s/he has to utilize the results (e.g., dimensions) of the previous layers. $IC$ and $FI$ of the first layer can be directly observed from the input image and global configuration of the accelerator. For the Wop-config of each Conv2D layer, the adversary could build a library of combinations of Wop-config and Opt-config, and then estimate their EM trace patterns. The candidate with high similarity to the observed EM trace of the target NN could be considered as the correct hypothesis with high confidence. %For example, the similarity can be measured by the Pearson correlation coefficient. 
%In practice, an attacker should cover the searching space of each Conv2D layer $SS_{Conv2D_i}$, where we assume an $NN$ have $l$ layers and $i$ represents the $i^{th}$ Conv2D layer. 
%Therefore, we can use the brute force attack searching space ($SS_{NN}$) to 
 %metric shown in Eq.~\ref{eq:bf}.
% \begin{equation}\label{eq:bf}
%     Sec_b = SS_{NN}
% \end{equation}

\subsubsection{Search space for individual Conv2D layers}\label{sec:bf_conv2d}
For a Conv2D layer, generally we assume $IC, FI$ are derived from the previous Conv2D layer's $OC, FO$, or the Pooling layer. Other parameters, including $\{K, FO, OC, fih_b, fiw_b, ic_b, oc_b, vt\}$, remain to be discovered. Some of these parameters follow some conventions that can be used as hints for guessing. \textbf{Hint 1:} The $K$ of the $1^{st}$ Conv2D layer might be 3, 5, or 7, and the $K$ of the rest Conv2D layers might be 1 or 3. \textbf{Hint 2:} $FO$ depends on the $FI$ and the stride of the kernel, which is normally 1 or 2. When $FI$ is smaller than 8, the stride will be 1. \textbf{Hint 3:} $OC$ depends on $IC$ and $BLOCK\_IN$, where $IC$ is expected as a multiple of $BLOCK\_IN$. An exception is that the $1^{st}$ Conv2D layer usually has an $IC$ smaller than $BLOCK\_IN$, so the VTA will convert it to $IC = BLOCK\_IN$ with dummy input channels. If representing the relationship between $OC$ and $IC$ as $OC = f_{oc} \times IC$, then $f_{oc} \in \{\frac{1}{4}, \frac{1}{2}, 1, 2, 4\}$. Note that $f_{oc}=\frac{1}{4}$ and $\frac{1}{2}$ do not happen when $IC = BLOCK\_IN$, and $f_{oc}=\frac{1}{4}$ do not occur when $IC = 2 \times BLOCK\_IN$.
Hence, the searching space ($SS$) for $K$, $FO$, $OC$ are
\vspace{-0.5em}
\begin{equation}\label{eq:wop-config_guess}
\small
\begin{array}{l}
SS_{K~} = \left\{ \begin{array}{cl}
3 &, i = 1 \\
2 &, i > 1
\end{array} \right. \\
SS_{FO} = \left\{ \begin{array}{cl}
1 &, FI < 8 \\%FI \bmod 2 = 0 \\
2 &, Otherwise
\end{array} \right. \\
SS_{OC} = \left\{ \begin{array}{cl}
3 &, IC = BLOCK\_IN \\
4 &, IC = 2 \times BLOCK\_IN \\
5 &, Otherwise
\end{array} \right. \\
\end{array}
\end{equation}

Following Eq.~\ref{eq:wop-config_guess}, an attacker can formulate the searching space for the potential Wop-conifgs of $Conv2D_{i}$. For a specific Wop-config, the size of its Opt-config searching space can be derived from Eq.~\ref{eq:opt-config_guess}, again we use the subscript $_b$ to represent the detailed value of each parameter out of many possibilities. 
\vspace{-0.5em}
\begin{equation}\label{eq:opt-config_guess}
\small
\left\{ \begin{array}{l}
SS_{fih_{b}} = SS_{fiw_{b}} = \left\lceil log_2\frac{FI}{4} \right\rceil \\
\\
SS_{ic_{b}} = \left\lceil log_2\frac{IC}{BLOCK\_IN} \right\rceil + 1\\
\\
SS_{oc_{b}|OC} = \left\lceil log_2\frac{OC}{BLOCK\_OUT} \right\rceil + 1\\
\\
SS_{vt} = \max(vt)
\end{array}\right.
\end{equation}

Assuming $SS_c = SS_K \times SS_{FO} \times SS_{fih_b} \times SS_{fiw_b} \times SS_{ic_b} \times SS_{vt}$, the searching space $SS_{Conv2D_{i}}$ of $Conv2D_{i}$ can be derived using Eq.~\ref{eq:ssconv2d}: 
\vspace{-0.5em}
\begin{equation}\label{eq:ssconv2d}
\small
SS_{Conv2D_i} =  SS_c \times \sum_{OC\in \Omega_{OC}} SS_{oc_{b}|OC}
\end{equation}
where $\Omega_{OC}$ denotes the corresponding values in the searching space of $OC$. For example, as shown in Tab.~\ref{tab:vgg19_bf}, the $2^{nd}$ Conv2D layer has $IC = 64$, thus $SS_{OC} = 5$ ($BLOCK\_IN$ and $BLOCK\_OUT$ are set as 16 in Sec. \ref{sec:vta-config}), and the specific values $\Omega_{OC} = \{16, 32, 64, 128, 256\}$ with possible $SS_{oc_{b}|OC}$ is 1, 2, 3, 4, 5, respectively. Considering the derivation $SS_c = 864$ for the $2^{nd}$ Conv2D layer in VGG-19, thus the searching space for this layer is $SS_{Conv2D_2} = 12960$.

\begin{table*}\label{}
\centering
\caption{VGG-19 under \textit{balance mode} with different scheduling}
\label{tab:vgg19_bf}
\begin{tabular}{c|ccccc|cc|cc} 
\hline
\begin{tabular}[c]{@{}c@{}}\textbf{e$_i$}\\\end{tabular} & \textbf{Conv2D$_i$} & \textbf{IC} & \textbf{FI} & \begin{tabular}[c]{@{}c@{}}\textbf{Independent}\\\textbf{$SS_{Conv2D_i}$}\end{tabular} & \begin{tabular}[c]{@{}c@{}}\textbf{Wop-config}\\\textbf{Ground Truth}\end{tabular} & \begin{tabular}[c]{@{}c@{}}\textbf{AutoTVM}\\\textbf{Opt-config}\end{tabular} & \begin{tabular}[c]{@{}c@{}}\textbf{AutoTVM}\\\textbf{}$SS_{Conv2D_i}$\\\end{tabular} & \begin{tabular}[c]{@{}c@{}}\textbf{\ourtit}\\\textbf{Opt-config}\end{tabular} & \begin{tabular}[c]{@{}c@{}}\textbf{\ourtit}\\\textbf{$SS_{Conv2D_i}$}\end{tabular} \\ 
\hline
1 & 1 & 16 & 224 & 2592 & {[}16, 64, 3, 224, 224] & {[}1, 2, 32, 4, 2] & 2592 & {[}1, 2, 32, 4, 2] & 2592 \\ 
\hline
2 & 2 & 64 & 224 & 12960 & {[}64, 64, 3, 224, 224] & {[}1, 1, 8, 8, 2] & 12960 & {[}1, 1, 8, 8, 2] & 12960 \\ 
\hline
3 & 3 & 64 & 112 & 9000 & {[}64, 128, 3, 112, 112] & {[}1, 2, 8, 4, 2] & 9000 & {[}1, 2, 8, 4, 2] & 9000 \\ 
\hline
4 & 4 & 128 & 112 & 16000 & {[}128, 128, 3, 112, 112] & {[}1, 2, 8, 4, 2] & 20 & {[}1, 1, 4, 4, 2] & 16000 \\ 
\hline
5 & 5 & 128 & 56 & 10240 & {[}128, 256, 3, 56, 56] & {[}1, 2, 4, 2, 2] & 10240 & {[}1, 2, 4, 2, 2] & 10240 \\ 
\hline
6 & \begin{tabular}[c]{@{}c@{}}6\\7\\8\end{tabular} & 256 & 56 & 16000 & {[}256, 256, 3, 56, 56] & {[}1, 2, 4, 2, 2] & \begin{tabular}[c]{@{}c@{}}20\\1\\1\end{tabular} & \begin{tabular}[c]{@{}c@{}}{[}1, 2, 8, 2, 2]\\{[}1, 2, 4, 2, 2]\\{[}1, 1, 8, 2, 2]\end{tabular} & \begin{tabular}[c]{@{}c@{}}16000\\16000\\16000\end{tabular} \\ 
\hline
7 & 9 & 256 & 28 & 9000 & {[}256, 512, 3, 28, 28] & {[}1, 1, 1, 1, 2] & 9000 & {[}1, 1, 1, 1, 2] & 9000 \\ 
\hline
8 & \begin{tabular}[c]{@{}c@{}}10\\11\\12\end{tabular} & 512 & 28 & 12960 & {[}512, 512, 3, 28, 28] & {[}1, 4, 4, 1, 2] & \begin{tabular}[c]{@{}c@{}}12960\\1\\1\end{tabular} & \begin{tabular}[c]{@{}c@{}}{[}1, 1, 4, 1, 2]\\{[}1, 2, 4, 1, 2]\\{[}1, 4, 4, 1, 2]\end{tabular} & \begin{tabular}[c]{@{}c@{}}12960\\12960\\12960\end{tabular} \\ 
\hline
9 & \begin{tabular}[c]{@{}c@{}}13\\14\\15\\16\end{tabular} & 512 & 14 & 5760 & {[}512, 512, 3, 14, 14] & {[}1, 4, 1, 1, 2] & \begin{tabular}[c]{@{}c@{}}5760\\1\\1\\1\end{tabular} & \begin{tabular}[c]{@{}c@{}}{[}1, 4, 1, 1, 2]\\{[}1, 2, 1, 1, 2]\\{[}1, 2, 1, 2, 2]\\{[}1, 1, 1, 2, 2]\end{tabular} & \begin{tabular}[c]{@{}c@{}}5760\\5760\\5760\\5760\end{tabular} \\ 
\hline
\multicolumn{6}{c|}{\textbf{SS$_{NN}$}} & \multicolumn{2}{c|}{ 62559} & \multicolumn{2}{c}{ 169712} \\ 
\hline
\multicolumn{6}{c|}{\textbf{All Conv2D Execution Time (ms)}} & \multicolumn{2}{c|}{ 1670.02} & \multicolumn{2}{c}{1721.19} \\
\hline
\end{tabular}
\vspace{-1.5em}
\end{table*}
%\vspace{-0.05em}

\subsubsection{Searching space for the entire neural network}\label{sec:bf_nn}

An attacker needs to iterate all possible Wop-configs and their Opt-configs, to compare the guessed EM leakage with the obtained NN EM leakage starting from the $1^{st}$ Conv2D layer. We can narrow the search space based on two facts: (1) Multiple Conv2D layers exist in one NN with the same Wop-config; (2) A specific Opt-config may be suitable for different Wop-configs to execute effectively. A strong and practical scenario is that an attacker only tries to utilize the prior knowledge before s/he really has to search the entire space. For example, an attacker is always applying the recovered Wop-config and Opt-config of the previous layer first. In other words, s/he has to checks all these already-discovered Wop-config ($SG_{wop}$) and Opt-config sets ($SG_{opt}$), only if the previous knowledge is not working. When both tests fail, s/he iterates other possible configurations. We show an example in Tab.~\ref{tab:vgg19_bf} column \textbf{AutoTVM Opt-config}, where $e_3$ and $e_4$ employ the same Opt-config. We calculate the searching space for each Conv2D layers and derive the $NN$ searching space $SS_{NN}$, which also represents the basic security level.
\vspace{-0.8em}
\begin{equation}\label{eq:ssnn}
\small
    SS_{NN} = \sum_{i=1}^{l}SS_{Conv2D_i}
\end{equation}

\subsubsection{EM obfuscation scheme}
For defense, one can use EM obfuscation. In details, a designer can follow Eq.~\ref{eq:pcgemm} to calculate the $OpC_{GEMM}$ in each Conv2D layer and find all potential EM obfuscation Wop-configs and Opt-config, using Eq.~\ref{eq:M} and~\ref{eq:N}. According the combination theory, the best choice of layers to apply EM obfuscation is $\left\lceil \frac{l}{2} \right\rceil$, where $l$ is the number total layers. Therefore, the designer can randomly select 8 layers out of the 16 layers of VGG-19 to obfuscate. Consequently, the searching space of brute force attack will be increased to a huge number, since the Conv2D configurations reflected in the EM trace does not help with reverse engineering at all. Further, as the Eq.~\ref{eq:Conv2D_EM} shows, $S$ can be changed (elongated), which will affect the similarity calculation in brute force attack. It is hard for an attacker to find a unique correct Wop-config for the current layer. If they ignore $S$ and only compare $M$, $N$, and $w_c$, it will induce many possible Wop-configs. 

\subsection{\ourtit~Framework}\label{sec:NNReArch}
% \vspace{-0.5em}
\begin{figure}[htpb!]
\centering
  \includegraphics[width=\linewidth]{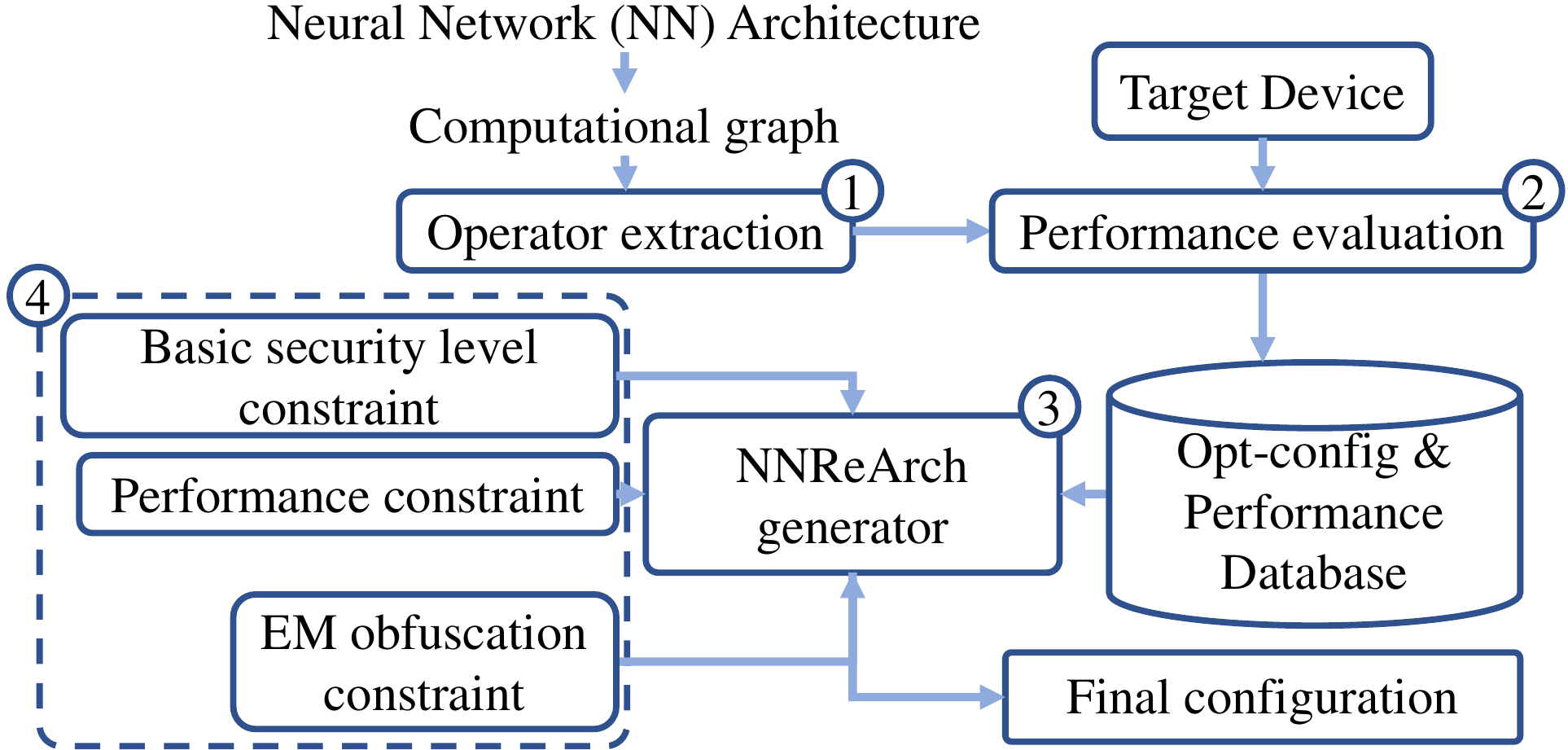}
  \caption{\ourtit~framework design overview.}
%   \Description{.}
  \label{fig:fram_view}
%   \vspace{-0.5em}
\end{figure}

\begin{table}[htpb!]
\centering
\caption{All possible workload configuration of Conv2D layers in VGG-19 applied in \ourtit~to build the Opt-config $\&$ Performance Database (Fig.~\ref{fig:fram_view}). }
\label{tab:dataset_build}
\begin{tabular}{c|ccc}
\hline
\textbf{e}$_i$ & \textbf{Wop-config} & \begin{tabular}[c]{@{}c@{}}\textbf{Ps$_{e_i}$}\\\textbf{size}\\\end{tabular} & \begin{tabular}[c]{@{}c@{}}\textbf{NO. of}\\\textbf{Usage}\end{tabular} \\ \hline
1 & {[}16, 64, 3, 224, 224] & 61 & 1 \\
2 & {[}64, 64, 3, 224, 224] & 183 & 1 \\
3 & {[}64, 128, 3, 112, 112] & 219 & 1 \\
4 & {[}128, 128, 3, 112, 112] & 292 & 1 \\
5 & {[}128, 256, 3, 56, 56] & 316 & 1 \\
6 & {[}256, 256, 3, 56, 56] & 395 & 3 \\
7 & {[}256, 512, 3, 28, 28] & 315 & 1 \\
8 & {[}512, 512, 3, 28, 28] & 378 & 3 \\
9 & {[}512, 512, 3, 14, 14] & 216 & 4 \\ \hline
\end{tabular}
\vspace{-1.5em}
\end{table}

% We next present our NN reconfiguration architecture, namely \ourtit, to protect the structural IP. 
\ourtit~is an automated tensor program generator that can mitigate the DNN-architecture-relevant EM side-channel leakage from the target device. Fig. \ref{fig:fram_view} shows the workflow of \ourtit. In step \textcircled{1}, we implement the operator extraction. The input is an abstracted DAG NN architecture, and the output is the operator expression list, as shown in  Tab.~\ref{tab:dataset_build} column $\textbf{e}_i$. Each of these operators is for certain Conv2D layers with different configurations. Through the TVM compiler we can generate different Opt-configs for a specific $e_i$ workload, as shown in the column \textbf{Ps}$_{e_i}$ \textbf{size}, the number of compilable Opt-configs ($Ps_{e_i}$). We evaluate their performance on the target device in step \textcircled{2}. We record all $Ps_{e_i}$ and their corresponding execution time in an ``Opt-config \& Performance Database''. Step \textcircled{3} denotes the proposed \ourtit, where the user may apply three constraints shown in \textcircled{4} to generate the corresponding low-level NN execution codes. In addition, they can also trade off the security and performance through these constraints. Two Opt-config selecting modes are provided for users: \textit{balance mode} and \textit{secure mode}. In \textit{balance mode}, \ourtit\ selects the high-performance Opt-config while ensuring the architecture confidentiality level. Specifically, the generator will first select the high-performance Opt-config. \textit{Secure mode} prioritizes security, by randomly choosing Opt-configs and a certain number of layers to apply the EM obfuscation, in order to significantly enlarging searching space. In Fig.~\ref{fig:fram_view}, the ``Basic security level constraint'' is related to maximizing the $SS_{NN}$ in both modes. The ``Performance constraint'' controls the \textit{balance mode} to satisfy the performance requirement, i.e., runtime overhead. The ``EM obfuscation constraint'' is customized by the user, who only needs to provide the number of layers to obfuscate, and \ourtit~ will generate low-level candidate code to satisfy all constraints. 

\section{Evaluation and Discussion}\label{sec:eval}

In this section, we evaluate the performance of \ourtit, and compare it with AutoTVM, using the most commonly utilized DNN architectures, including VGG-16, VGG-19, ResNet-18, and ResNet-34. For sake of clarity, we list all possible workload configuration (i.e., ground truth) of the Conv2D layers of VGG-19 in Tab.~\ref{tab:dataset_build}. For example, ``$e_{i} = 1$” represents the $1^{st}$ type of Conv2D layer, and its corresponding ``NO. of usage = 1'' means that this layer type is only used once in VGG-19. From the brute force attack perspective, if given the correct $IC$ and $FI$, the search space $SS_{Conv2D_1}$ of $Conv2D\_1$ is 2592. Therefore, the attacker will have to iterate entire $SS_{Conv2D_1}$, to find out the best matching workload. 

We present the technical detail of deploying \ourtit~on VGG-19 as an example, and illustrate the experimental results of all other DNN architectures in Fig.~\ref{fig:perf_ev}.

\subsection{Applying \ourtit~on VGG-19}
We firstly apply the \textit{balance mode}  of \ourtit~on VGG-19, which considers the tradeoff between security and performance, and the results are shown in Tab.~\ref{tab:vgg19_bf}. Specifically, we apply the security constraint as ``maximizing the $Sec_{b}$'' and the performance constraint as ``shortest execution time''. Note that this evaluation does not include the EM obfuscation. The performance of applying \ourtit~is mainly shown in column \textbf{NNReArch Opt-config} in Tab.~\ref{tab:vgg19_bf}. The total execution time of all VGG-19 Conv2D layers is 1721.19ms. Compared with the original performance using ``AutoTVM'' scheduling (column \textbf{AutoTVM Opt-config} in Tab.~\ref{tab:vgg19_bf}), the deployment of \ourtit~only incurs $3.06\%$ performance overhead. In contrast, the searching space ($SS_{Conv2D_i}$) is significantly increased, with $SS_{NN} = 169712$. Thus, in terms of the \textit{balance mode}, applying \ourtit~(w/o EM obfuscation) increases the difficulty of DNN architecture extraction for about 2.71 times.

\begin{table}
\centering
\caption{\ourtit~applied on 8 Conv2D layers with \textit{secure mode}  (EM obfuscation)}
\label{tab:nnreach_w_EM_vgg19}
\scalebox{1}{
\begin{tabular}{c|ccc} 
\hline
\textbf{e$_i$} & \textbf{Conv2D$_i$} & \begin{tabular}[c]{@{}c@{}}\textbf{\ourtit~}%\\\textbf{(w/ Emob.)}
\\\textbf{Wop-config}\end{tabular} & \begin{tabular}[c]{@{}c@{}}\textbf{\ourtit}%\\\textbf{(w/ Emob.)}
\\\textbf{Opt-config}\end{tabular} \\ 
\hline
1 & 1 & {[}16, 64, 3, 224, 224] & {[}1, 2, 32, 4, 2] \\ 
\hline
2 & 2 & {[}64, 64, 3, 224, 224] & {[}1, 1, 8, 8, 2] \\ 
\hline
3 & 3 & {[}64, 128, 3, 112, 112] & {[}1, 2, 8, 4, 2] \\ 
\hline
4 & 4 & {[}128, 128, 3, 112, 112] & {[}1, 1, 4, 4, 2] \\ 
\hline
5 & 5 & {[}128, 256, 3, 56, 56] & {[}1, 2, 4, 2, 2] \\ 
\hline
6 & \begin{tabular}[c]{@{}c@{}}6\\7\\8\end{tabular} & \begin{tabular}[c]{@{}c@{}}{[}256, 64, 3, 56, 112]$^1$\\{[}64, 256, 3, 112, 112]$^2$\\{[}256, 256, 3, 112, 56]$^3$\end{tabular} & \begin{tabular}[c]{@{}c@{}}{[}1, 2, 8, 2, 2]\\{[}1, 2, 4, 2, 2]\\{[}1, 1, 8, 2, 2]\end{tabular} \\ 
\hline
7 & 9 & {[}256, 512, 3, 28, 28] & {[}1, 1, 1, 1, 2] \\ 
\hline
8 & \begin{tabular}[c]{@{}c@{}}10\\11\\12\end{tabular} & \begin{tabular}[c]{@{}c@{}}{[}512, 512, 3, 28, 28]\\{[}512, 512, 3, 28, 28]\\{[}512, 2048, 3, 28, 14]$^4$\end{tabular} & \begin{tabular}[c]{@{}c@{}}{[}1, 1, 4, 1, 2]\\{[}1, 2, 4, 1, 2]\\{[}1, 4, 4, 1, 2]\end{tabular} \\ 
\hline
9 & \begin{tabular}[c]{@{}c@{}}13\\14\\15\\16\end{tabular} & \begin{tabular}[c]{@{}c@{}}{[}2048, 512, 3, 7, 7]$^5$\\{[}512, 2048, 3, 7, 7]$^6$\\{[}2048, 512, 3, 7, 7]$^7$\\{[}512, 512, 3, 7, 14]$^8$\end{tabular} & \begin{tabular}[c]{@{}c@{}}{[}1, 4, 1, 1, 2]\\{[}1, 2, 1, 1, 2]\\{[}1, 2, 1, 2, 2]\\{[}1, 1, 1, 2, 2]\end{tabular} \\ 
\hline
\multicolumn{3}{c|}{\textbf{All Conv2D Execution Time (ms)}} & 1927.59 \\
\hline
\end{tabular}}
\vspace{-1.5em}
\end{table}

We further apply the \textit{secure mode}  of \ourtit~(i.e., with EM obfuscation) to prioritize the DNN model architecture confidentiality. Specifically, we select 8 layers of VGG-19 to schedule their tensor program, making them to have the same EM trace characteristics. The Opt-config setting of the obfuscated workload are listed in Tab.~\ref{tab:nnreach_w_EM_vgg19}, and here we use superscripts ({1,2,...,8}) to annotate the obfuscated layers. Compared with the ground truth configuration (column \textbf{
Wop-config Ground Truth} in Tab.~\ref{tab:vgg19_bf}), the obfuscated EM traces will lead the reverse engineering attack to wrong workload. For example, the EM obfuscation breaks the performing divergence of the EM leakage on different convolution layers, when the attacker reasons each Conv2D layer, i.e., s/he will derive wrong workloads that negatively affect the guess on followed layers, disturbing the consistency of adjacent Conv2D layers. As a result, when the exteriors of different Wop-configs have the same patterns, attackers have to handle a huge amount of puzzles and guesses. Thus the searching space and attacking effort will increase massively.

As discussed in Sec.~\ref{sec:EMob}, the EM obfuscation needs to add pause opcodes to obfuscate the stall time, to make the obfuscated layer performing almost the same as the ``imitated layer'', which may brings performance loss. In our experimental evaluation on VGG-19 shown in Tab.~\ref{tab:nnreach_w_EM_vgg19}, the obfuscated DNN model consumes 1927.59 ms to execute all Conv2D layers, which has an approximate $15.42\%$ performance overhead compared with the AutoTVM.

\subsection{Discussion: security and performance trade-off}

\begin{figure}[htpb!]
     \centering
          \begin{subfigure}[b]{0.8\linewidth}
         \centering
         \includegraphics[width=\linewidth]{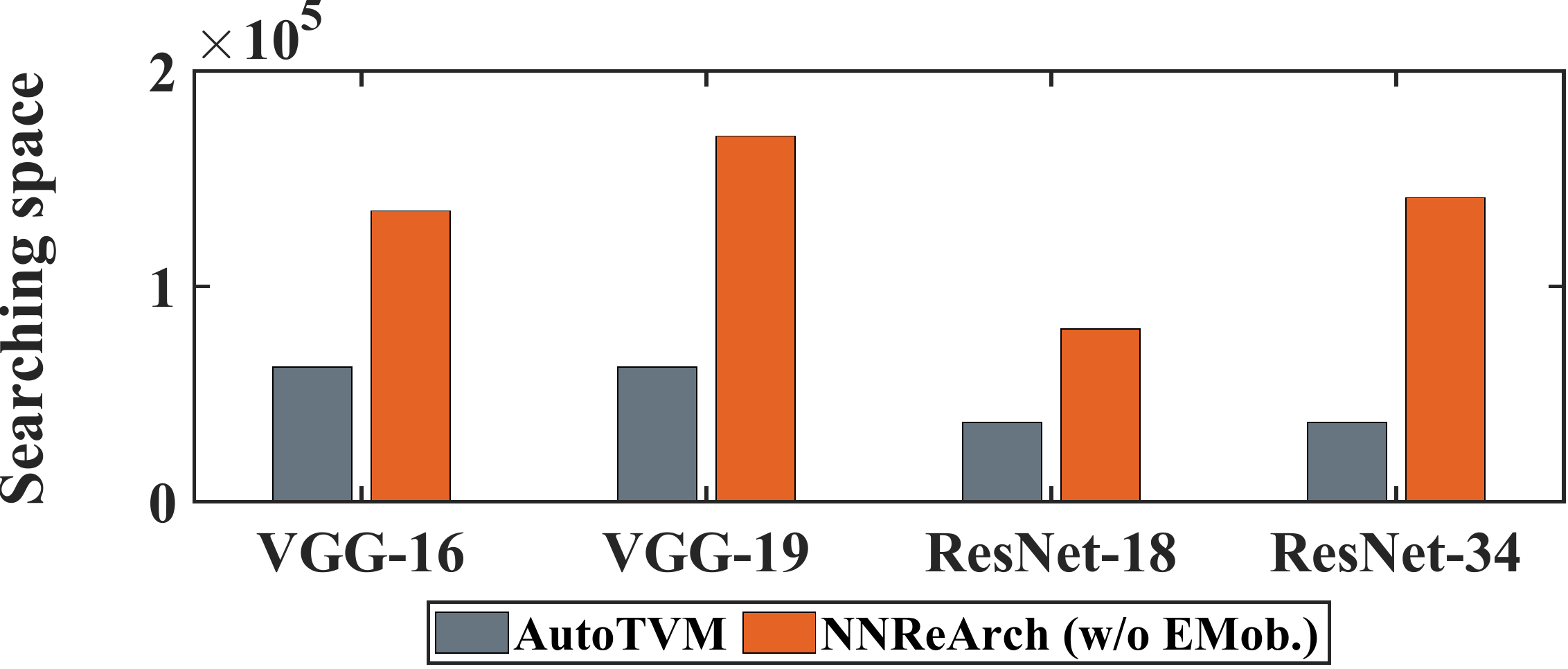}
         \caption{Security level measurement}
         \label{fig:AvsN_secb}
     \end{subfigure}
     \begin{subfigure}[b]{0.8\linewidth}
         \centering
         \includegraphics[width=\linewidth]{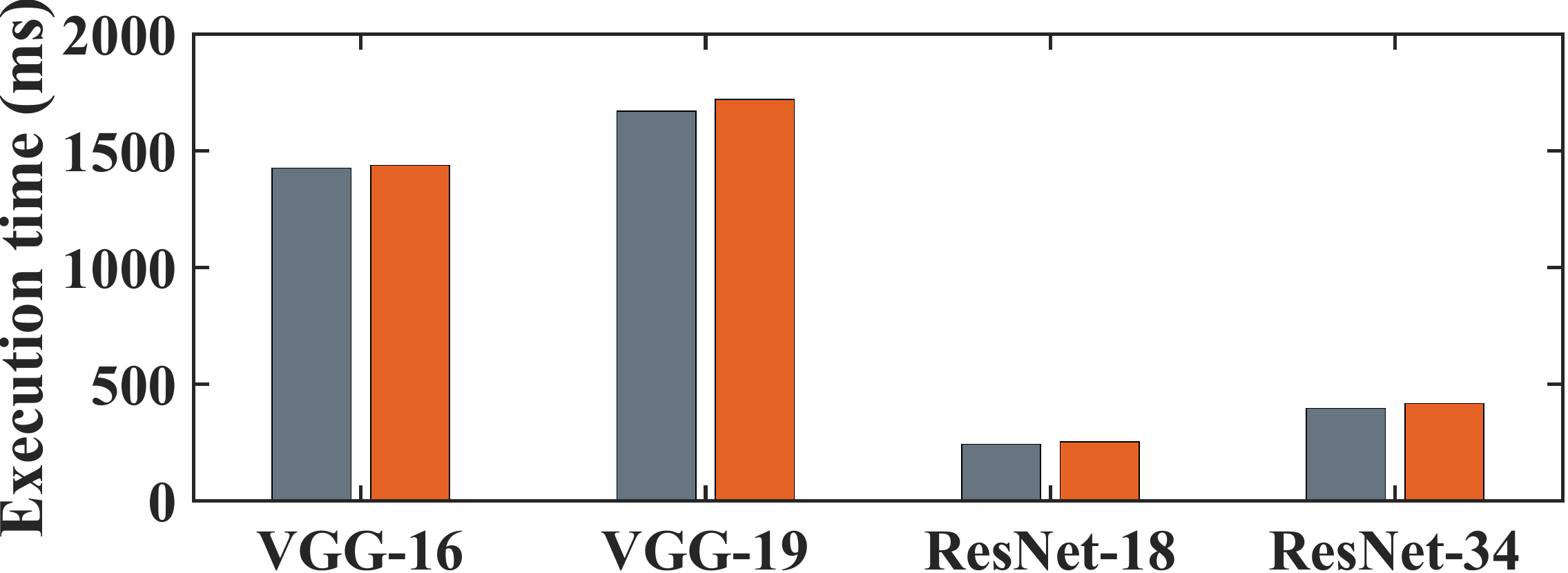}
         \caption{Performance measurement}
         \label{fig:AvsN_perform}
     \end{subfigure}

        \caption{Performance evaluation and comparison between AutoTVM and the \ourtit~\textit{balance mode} of \ourtit.}
        \label{fig:perf_ev}
        \vspace{-1.6em}
\end{figure}

Since most existing DNN development frameworks (including both open-source and commercial) still target at performance, therefore, we evaluate the \textit{balance mode} of \ourtit~on the popular DNN architectures from VGG and ResNet to draw generic conclusions. We first compare the searching space  ($SS_{NN}$) of DNNs from the same family. For example, VGG-16 and VGG-19 are constructed by the same Conv2D layer types, but only with different number of layers of the same Wop-config. Therefore, the attacking difficulty for DNNs in the same family is almost equal using AutoTVM. As affirmed in Fig.~\ref{fig:AvsN_secb}, the searching space is the same for VGG-16 and VGG-19, as well as for ResNet-18 and ResNet-34. This indicates a vulnerability of these performance-only optimization frameworks, i.e., using a larger DNN model does not make it more challenging to reverse engineer the model architecture. In contrast, the proposed \ourtit~framework constructively leverages the model size to maximizes the searching space of each Conv2D layer, making the architecture of deeper DNNs more secure, as shown in Fig.~\ref{fig:AvsN_secb}. Fig.~\ref{fig:AvsN_perform} illustrates the performance comparison between AutoTVM and \ourtit, which demonstrates that \ourtit~only incurs trivial execution time overhead on all evaluated DNN architectures compared with AutoTVM.

For the \textit{secure mode}  of \ourtit, we demonstrate that a carefully crafted DNN model (e.g., the one in Tab.~\ref{tab:nnreach_w_EM_vgg19}), can significantly challenge the model extraction attacks with relatively higher performance loss (15\%). Moreover, the \textit{secure mode}  enables the designer to either randomly choose Conv2D layers for EM obfuscation, or apply unexpected stall time to break the association between the EM side-channel leakage and workload configuration, 
thus providing more flexibility to the DNN model security enhancement. 

\section{Conclusion}\label{sec:discussion}
In this paper, we study the association between DNN model architecture configuration and EM side-channel leakage. Using an open-source deep-learning accelerator VTA as our experimental platform, we discover the low-level code causes of the EM side-channel-enabled DNN architecture reverse engineering attacks. Furthermore, we present \ourtit, a DNN model architecture defense framework against side-channel attacks. Enabling flexible DNN configuration between performance and security, \ourtit~integrates two modes, \textit{balance mode} that targets at increasing the searching space of DNN model architectures, and \textit{secure mode} that employs EM obfuscation to cancel the difference between different model layers. Different from the existing solutions, the proposed framework is built on popular open-source DNN compilation tools, VTA, making it a generic defense method. 

\bibliographystyle{IEEEtran}
\bibliography{FCCM2022}
\end{document}